\documentclass[12pt]{article}
\pdfoutput=1

\usepackage{amsmath,amssymb, cite,multirow}   
\usepackage{epsfig}                    
\usepackage[matrix,arrow]{xy}          
\usepackage{xspace,bbold}              
\usepackage{stmaryrd}                  
\usepackage[colorlinks=true,linkcolor=blue,citecolor=blue,linktoc=page]{hyperref}

\numberwithin{equation}{section}       

\allowdisplaybreaks                    

\newcommand{\beq}{\begin{equation}}
\newcommand{\eeq}{\end{equation}}
\newcommand{\baed}{\begin{aligned}}
\newcommand{\eaed}{\end{aligned}}
\newcommand{\bea}{\begin{eqnarray}}
\newcommand{\eea}{\end{eqnarray}}

\newcommand{\dd}{\mathrm{d}}
\newcommand{\ee}{\mathrm{e}}
\newcommand{\ii}{\mathrm{i}}

\newcommand{\del}{\partial}

\newcommand{\sla}{\slash\!\!\!\!}

\newcommand{\bbR}{\mathbb{R}}

\DeclareMathOperator{\SU}{\mathit{SU}}

\DeclareMathOperator{\GL}{\mathit{GL}}

\DeclareMathOperator{\Spin}{\mathit{Spin}}

\DeclareMathOperator{\coker}{coker}
\DeclareMathOperator{\adj}{ad}

\newcommand{\id}{\mathbf{1}}

\DeclareMathOperator{\tr}{tr}

\DeclareMathOperator{\vol}{vol}

\newcommand{\Gs}[1]{\Gamma(#1)}
\newcommand{\GM}[2]{\big<#1,#2\big>}

\newcommand{\comm}[2]{\left[#1,#2\right]}
\newcommand{\met}[2]{\big<#1,#2\big>}
\newcommand{\Met}[2]{\left(#1,#2\right)}
\newcommand{\norm}[1]{\left|#1\right|}

\newcommand{\der}{\partial}

\newcommand{\Lgen}{\mathbb{L}}
\newcommand{\Bgen}[2]{\left\llbracket#1,#2\right\rrbracket}

\newcommand{\Edd}{\mathit{E}_{d(d)}}

\newcommand\e{\mathrm{e}}

\newcommand{\ba}{\begin{eqnarray*}}
\newcommand{\ea}{\end{eqnarray*}}
\newcommand{\ban}{\begin{eqnarray}}
\newcommand{\ean}{\end{eqnarray}}

\newcommand{\Tint}{\hat{T}}

\begin{document}

\begin{titlepage}

\begin{center}

\rightline{\small CERN-PH-TH/2014-138}
\rightline{\small Imperial/TP/2014/DW/03}
\rightline{\small IPhT-t14/105}
\rightline{\small ITP-UH-12/14}

\vskip 1.0cm

{\Large \bf Generalised geometry for string corrections}

\vskip 1.5cm

{\bf Andr\'e Coimbra},$^a$ {\bf Ruben Minasian},$^{b}$ {\bf Hagen Triendl}$^{c}$ and {\bf Daniel Waldram}$^{d}$\\

\vskip 0.5cm

{}$^{a}${\em Institut f\"ur Theoretische Physik \& \\
  Center for Quantum Engineering and Spacetime Research,\\
Leibniz Universit\"at Hannover, Appelstra{\ss}e 2, 30167 Hannover, Germany }

\vskip 0.4cm

{}$^{b}${\em Institut de Physique Th\'eorique, CEA Saclay,\\
Orme de Merisiers, F-91191 Gif-sur-Yvette, France}

\vskip 0.4cm

{}$^{c}${\em Theory Division, Physics Department, CERN,\\
CH-1211 Geneva 23, Switzerland}

\vskip 0.4cm

{}$^{d}${\small \em Department of Physics, Imperial College London, \\
Prince Consort Road, London, SW7 2AZ, UK}

\vskip 0.7cm

{\tt andre.coimbra@itp.uni-hannover.de, ruben.minasian@cea.fr, hagen.triendl@cern.ch, d.waldram@imperial.ac.uk} \\

\end{center}

\vskip 1.0cm

\begin{center} {\bf ABSTRACT } \end{center}

We present a general formalism for incorporating the string corrections in generalised geometry, which  necessitates the extension of the generalised tangent bundle. Not only are such extensions obstructed, string symmetries and the existence of a well-defined effective action require a precise choice of the (generalised) connection. The action takes a universal form given by a generalised Lichnerowitz--Bismut theorem. As examples of this construction we discuss the corrections linear in $\alpha'$ in heterotic strings and the absence of such corrections for type II theories.

\vfill

\noindent\today


\end{titlepage}


\tableofcontents


\section{Introduction}


Generalised complex geometry~\cite{Hitchin,Gualtieri} provides a geometric description of string theory that makes its symmetries manifest. The generalised tangent bundle incorporates information about the space-time geometry and the topology of the $B$-field and so is well-suited to describe the on-shell supersymmetry~\cite{GMPT} and give a handle on understanding non-local symmetries such as T-duality~\cite{GMPW}. In fact generalised geometry provides a compact description of the full off-shell type II supergravity theories using a generalised version of the Levi--Civita connection that encodes the geometry of NSNS fields. The geometrical description of the bosonic theory was first considered in~\cite{Siegel,HK,JLP} using the language of Double Field Theory (DFT)~\cite{dft}. The full theory with fermions was given using generalised geometry in~\cite{CSW1} (see also~\cite{Jeon:2012hp} for a subsequent DFT version, given to all orders in fermions). Suitable generalisations of the basic construction, involving the exceptional groups $\Edd$, describe all massless fields in type II string theory and M-theory in $d-1$ and $d$ dimensions respectively~\cite{Hull:2007zu,Pacheco:2008ps}. The dynamics are again encoded by the corresponding generalised Levi--Civita connections~\cite{CSW-11d}. That this geometry also describes the full ten- or eleven-dimensional theory was recently shown in~\cite{GGHNS} using the ideas of~\cite{dWN} and~\cite{EFT}. So far the generalised geometrical constructions have been classical.

In this paper we shall address two natural questions concerning the extensions of the formalism. The first question is whether it is possible to incorporate the $\alpha'$ expansion in the framework of generalised geometry.\footnote{We shall not speak about the string loops and the $g_s$ expansion here.} If so, does the result agree with the known string theoretic corrections? The second question may be thought of as a particular case of the first. Is there a generalised geometric description of the heterotic strings? The problem here is that the closure of the three-form $H$ which, as will be reviewed in sec.~\ref{sec:GG}, is essential for the standard construction of  \cite{Hitchin,Gualtieri} is no longer satisfied.\footnote{In type II theories we can think of $H$ as a curvature and of the $B$-field as a connection on a gerbe. For the heterotic string this is not longer true and leads to a number of subtleties explored in  particular in~\cite{Witten:1999eg}.}

Our conclusion will be that to correctly describe corrections generically one needs to extend the generalised tangent space. There can be important topological restrictions arising from the construction of the extended tangent space $E$. The underlying supergravity degrees of freedom and dynamics arise from choosing the appropriate sub-structures on $E$ and considering compatible connections. The action takes a universal form which arises as a generalised Lichnerowitz--Bismut theorem~\cite{BisTor}. A key new point here is that we naturally obtain the correct gravitational connection in the Bianchi identity. The $\alpha'$-corrected generalised geometry in the heterotic string context has previously been considered in~\cite{GF,Baraglia}. A similar construction describing heterotic $\alpha'$-corrections in the DFT formalism was presented in recent work~\cite{Bedoya:2014pma}. These results can also be compared with~\cite{Hohm:2013jaa}, which takes a different approach in the context of DFT by perturbing the generalised Lie derivative and the $O(n,n)$ metric to include higher-derivative corrections. However, in the latter approach the extra terms are not manifestly diffeomorphism invariant and, in addition, different generalised tensors receive corrections at different orders in $\alpha'$. As we discuss in section~\ref{sec:qb!}, the correct geometrical framework to understand such corrections is as lifts into the generalised tangent spaces described here.\footnote{While this paper was being completed a discussion of further corrections to the generalised Lie derivative for bosonic string theory appeared~\cite{HZ-new}, now also depending on the dynamical fields. We will not comment on the relation to the bosonic theory here.}

The heterotic Bianchi identity provides a natural starting point. It reads schematically
\begin{equation} \label{eq:hetBianchi}
\dd H = \tfrac{1}{4 } \alpha' (\tr F\wedge F -  \tr R\wedge R) \ ,
\end{equation}
where the first term is built out of the curvature two-form $R$ and the second of the gauge field strength $F$
of either the gauge group $G = E_8 \times E_8$ or $G= SO(32)$.  The word schematic refers to the fact that we have not specified yet what connections are used in \eqref{eq:hetBianchi}. A change of connection will result in a shift on the right-hand side by an exact form. In a number of applications,  e.g.\ when determining the consistency of heterotic compactifications,  one is interested in the integrated form of \eqref{eq:hetBianchi}, where the choice of connection is immaterial. On the other hand, the manifest $(0,1)$ world-sheet supersymmerty together with the choice of a hermitian covariant metric in space-time singles out  the (torsionful) connection
\begin{equation}\label{eq:Omegaminus}
 \Omega^{-} = \omega^{{\rm LC}} - \tfrac12 H \ .
\end{equation}
Here $\omega^{{\rm LC}}$ is the Levi-Civita connection and $H$ is understood as the $\mathfrak{o}(n)$-valued one-form $H_{ab} = \iota_{\hat e^a} \iota_{\hat e^b} H$. A priori, there is a freedom in choosing the sign of $H$, but this choice is correlated with the signs in the supersymmetry transformations.

The two terms in \eqref{eq:hetBianchi} present one with different types of problems. The gauge part comes with new degrees of freedom and an enlarged local symmetry. So it is clear from the beginning that new ingredients need to be added. For an Abelian gauge group the situation is well-understood. As we shall review in detail, by considering a backgrounds with a set of commuting isometries and dimensional reduction of an $SO(n+d,n+d)$ structure, one gets the Abelian version of~\eqref{eq:hetBianchi} with two set of $U(1)^d$ fields.\footnote{We shall denote the dimension of the  space-time by $n$ (typically $n=10$ or rather $n=(1,9)$). When we talk about the backgrounds with isometries or dimensional reductions the number of compact dimensions will be denoted by $d$.} More generally one can reduce on a general group manifold and obtain the non-Abelian versions of~\eqref{eq:hetBianchi}. A closely related notion, $B_n$ generalised geometry on a bundle $TM  \oplus \mathbb{R}\oplus T^*M$  is discussed in~\cite{Hitchin:Bn,Bn,Rubio}.  Incorporation of  the non-Abelian degrees of freedom requires a new extension of the generalised tangent bundle. Locally $E\simeq TM\oplus \adj{P_G} \oplus T^*M$, where $\adj{P_G}$ is the vector bundle for the adjoint representation of a Lie group $G$, that is, with fibres that are the Lie algebra $\mathfrak{g}$ of $G$. Such structures are known in the mathematics literature as transitive Courant algebroids~\cite{severa,csx}, and can always be constructed as dimensional reductions of conventional generalised geometry on a group manifold.\footnote{Heterotic flux (torsionful) backgrounds  provide a natural context for applying these constructions \cite{ags, xos} The role of  torsionful connections \eqref{eq:Omegaminus} in heterotic backgrounds has been discussed in \cite{mms}.}. We refer to section \ref{sec:nonAb} for explanations and definitions.  These structures were used in~\cite{GF,Baraglia} to construct heterotic actions. On the physics side, constructing heterotic theories by reduction has been discussed in~\cite{dnpw,David}, the form of the generalised Lie derivative and the local generalised geometry in the DFT context is given in~\cite{DFT-het}, and in~\cite{GM} it is shown how such generalised structures via dimensional reduction. There are global constraints associated with the extensions of the generalised tangent bundle (to be reviewed shortly) but as far as the consistency of the construction goes there are no restrictions on the choice of connection in the $\frac{1}{4} \alpha'  \tr F\wedge F$ factor.\footnote{As will be discussed in subsection \ref{sec:redgauge} the connection may be restricted by extra requirements, such as T-duality covariance.}

On the contrary, the gravitational contribution to \eqref{eq:hetBianchi} does not enlarge the symmetry group. It is tempting and in many ways natural to consider $GL(n,\mathbb{R})$ as a special case of the gauge group (see e.g. \cite{dnpw}), so that together with the gauge part one is considering a group $G\times\GL(n,\bbR)$. By doing so one can correctly reproduce the heterotic Bianchi identity. Such an approach was used to reproduce the heterotic dynamics using generalised geometry in~\cite{GF,Baraglia}. The problem however is that the new extended generalised tangent bundle is too large and there will be unphysical degrees of freedom. In particular, the $\GL(n,\bbR)$ connection in~\cite{GF,Baraglia} has to be put in by hand. This is reflecting the fact that the $\GL(n,\bbR)$ symmetry is not independent but is the group that rotates bases for the tangent and cotangent spaces. The solution is that when defining the dynamical degrees of freedom we specify a more refined structure. In particular we identify an $O(n)$ sub-bundle of $\GL(n,\bbR)$ and then identify this $O(n)$ structure with one of the $O(n)$ structures defined on the $TM\oplus T^*M$ part of the tangent space. Thus the overall structure group is simply $O(n)\times G \times O(n) $. We then require that the generalised connections which define the dynamics are compatible with the refined structure. We find we must slightly relax the constraint that the generalised connection is torsion-free, and demand instead that all the connections that appear in supergravity equations be unique. Remarkably, this procedure by construction gives the full Bianchi identity \eqref{eq:hetBianchi} with precisely the connection \eqref{eq:Omegaminus}. The generalised geometric construction moreover reproduces the known fact that the supersymmetry transformation of the gravitino should contain a covariant derivative using $ \Omega^{+} = \omega^{\rm LC} + \tfrac12 H$. All this is in accordance with the known string-theoretic calculations, but the discussion of the way the local data enters in the construction of the extended generalised tangent bundle is new.

Finally, we may turn to the effective actions. As shown in \cite{Siegel,HK,CSW1} even though a generalised Riemann tensor is not unambiguously defined by vanishing (generalised) torsion conditions, the generalised Ricci tensor is. Moreover, the Ricci scalar gives the full NSNS part of the effective action. The Ricci scalar can be constructed using the action of the generalised connection on spinor representations~\cite{CSW1}, in a way that is closely related to a generalisation of Lichnerowitz theorem due to Bismut~\cite{BisTor}, which states that there exists a pair of first-order operators, a covariant derivative and a Dirac operator respectively,\footnote{Note that the Dirac operator here is {\sl not} the trace of the covariant derivative. In general, the torsion parts in the respective operators may involve different physical degrees of freedom.} such that the difference of their squares is tensorial. The construction of the extended  generalised bundle provides us with the needed first-order operators (corrected up to order $\sim \alpha'$).  The order $\alpha'$ effective action $S$ is schematically written as
\begin{equation}
\label{eq:newBismut*}
     ( D^aD_a + D^{\alpha}D_{\alpha} - (\sla D)^2) \epsilon
          =  \tfrac14 S \epsilon + \gamma^{abcd} I_{abcd} \epsilon \ .
\end{equation}
It will not come as a surprise that $\sla D$, $D_a$ ($a=1, \cdots, n$) and $D_{\alpha}$ ($\alpha = 1, \cdots, \mbox{dim}(G)$) are the operators appearing in the supersymmetry variations of the dilatino, gravitino and gaugini respectively. Naively the right-hand side of \eqref{eq:newBismut*} contains not only a non-trivial scalar $S$ but also four-form $I$. However the latter is equal to the Bianchi identity and thus vanishes. In section~\ref{sec:higher} we show that after adding an $\GL(n,\bbR)$ factor and identifying the $O(n)\times G\times O(n)$ structure, we naturally find a composite fermion that corresponds to the gravitino curvature.
Making the variations of the composite field compatible with the variations of the gravitino while preserving the condition that $I$ above vanishes actually induces a whole hierarchy of higher order $\alpha'$ corrections. This is the generalised geometric version of analysis of~\cite{BdR}.

The appearance of a generalised Lichnerowitz--Bismut theorem is not an accident. As we discuss in section~\ref{sec:gen}, it is a generic consequence of supersymmetry. We show that one can construct a general formalism by which generalised geometry can be viewed as an infinite-dimensional version of the embedding tensor formalism~\cite{embed}. The generalised Lichnerowitz--Bismut theorem is then simply a result of a supersymmetric Ward identity~\cite{s-ward,embed-lecture}.

Interestingly, if one tries to apply this same set of ideas to type II theories (this suggestion goes back to \cite{dnpw}) and consider an extended generalised tangent bundle with $G=O(n)\times O(n)$ and then try and identify these with the $O(d)_+$ and $O(d)_-$ acting on $TM\oplus T^*M$, one finds one cannot consistently choose a generalised connection compatible with $O(n) \times O(n)$.  As will be explained in detail in Appendix~\ref{sec:typeII} depending on the allowed choices, one either ends up with ambiguities in the effective action or in the Bianchi identity. Hence $\dd H=0$ is the only consistent choice. The absence of corrections linear in $\alpha'$ in theories with 32 supercharges is well known, but the exercise is instructive and shows that the outlined procedure is restrictive.

In a slightly different direction, one can take the extended generalised generalised tangent space $E\simeq TM\oplus \adj{P_G} \oplus \adj{P_{\GL(n,\bbR)}}\oplus T^*M$, where we now take $P_{\GL(n,\bbR)}$ to be the frame bundle and try to restrict the symmetry parameters to elements of $TM \oplus \adj{P_G}\oplus T^*M$, that is diffeomorphisms, $B$-field gauge transformations and gauge transformations of the $G$-connections. To do so we need to determine a $\GL(n,\bbR)$ transformation in terms of a diffeomorphism. This requires a lift of vectors, essentially defining the $\GL(n,\bbR)$ transformation via
\begin{equation} \label{eq:relphys}
{\cal L}_v e^a = \tilde \Lambda^a{}_b e^b \ ,
\end{equation}
where $v \in \Gs{TM}$ is the generator of a diffeomorphisms, $\tilde \Lambda \in \Gs{\adj{P_{\GL(n,\bbR)}}}$ and $e^a$ is an arbitrary section of the frame bundle.\footnote{Changing the choice of section $e^a$ corresponds to a gauge transformation in $\tilde G$. Therefore, the choice of $e^a$ fixes a gauge in $\tilde G$.} This states that the $\GL(n,\bbR)$ transformations are generated by the action of diffeomorphisms on the auxiliary frame $e^a$. Note that due to the appearance of the Lie derivative ${\cal L}_v$, \eqref{eq:relphys} does not define a subbundle of the generalised tangent bundle. Rather, it defines a subspace in the space of sections on the original bundle $E$. In section~\ref{sec:qb!} we show that inserting \eqref{eq:relphys} into the natural metric on $E$ and the corresponding generalised Lie derivative leads to terms that look like higher-derivative corrections to the conventional expressions in the unextended case where $E\simeq TM\oplus T^*M$. In fact, when restricted to coordinate frames (and hence written in non-manifestly covariant form), these corrections match the $\alpha'$ corrections recently computed in DFT~\cite{Hohm:2013jaa}. However, in our context, viewing them as corrections is somewhat misleading, and not only because manifest diffeomorphism invariance is then lost. Globally the generalised vectors are sections of the extended generalised tangent space and not the conventional one: the only way to make sense of these expression is in terms of the extended space. Note that related discussions to this point recently appeared in~\cite{Bedoya:2014pma,HZ-het}.

As we see the complete construction requires both  global and local considerations. As mentioned, some of the former have appeared in the maths literature, in the context of transitive Courant algebroids, as have aspects of the latter, though without giving a natural explanation for the appearance of $\Omega^-$ in the Bianchi identity. The extension of the construction needed to correctly cover the gravitational contributions to the Bianchi identity and the important appearance of the local data is new. Our results for the effective action  reproduce the  known $\alpha'$-corrected supersymmetric actions, but the ``all-at-once" construction, based on the generalised Lichnerowitz-Bismut theorem, presented here is also original.

So far we have mostly concentrated on the heterotic Bianchi identity, with type II just providing an illustrative example of how a complete construction may rule out certain corrections. However we think that  the main principles/ingredients should apply in a more general discussion as well:

 \vskip 0.3cm
 \noindent {\bf Extensions $\,\,\,$} The first step is always to extend the generalised tangent bundle. We have discussed here the most intuitive extension using the local gauge symmetries of the theory. However one might expect generically a series of extensions, perhaps most naturally mirroring the local symmetries of the hierarchy of massive states in the corresponding string theory. The extended space always admits a conventional generalised Lie derivative that is first-order in the derivative. In the general formalism this can be regarded as identifying the infinite-dimensional analogue of the embedding tensor. These objects are then the main building blocks of the construction. They in particular determine the possible corrections to the Courant bracket.

  \vskip 0.3cm
 \noindent {\bf Obstructions $\,\,\,$} The extensions are obstructed. The triviality of the {\sl total} (gauge and tangent bundle) Pontryagin class is well known and has been much used in heterotic compactifications. Yet crucially, $H$ is not a curvature of a gerbe. In the type II case all possible extensions preserve the gerbe nature of the $B$-field, and eventually result in the absence of corrections linear in $\alpha'$. This happens due to subtle cancellations and there is no reason that further (e.g. $(\alpha')^3$) corrections should not appear.

  \vskip 0.3cm
 \noindent {\bf Connections $\,\,\,$} One may allow for the possibility that the extended generalised tangent bundle may have more symmetry than needed. It is important that the connections are compatible with the correct symmetries. This compatibility may in turn impose restrictions.

  \vskip 0.3cm
 \noindent {\bf Local actions $\,\,\,$} The generalised Lichnerowitz-Bismut theorem allows the computation of effective actions starting from the extended generalised tangent bundle.  The procedure is somewhat different from computing supersymmetric completions.  It is important to note that a particular correction to a given order in $\alpha'$ may induce all order corrections in the effective action (as is the case with the heterotic Bianchi identity \cite{BdR}). On the other hand, a family of  corrections (a priori all orders in $\alpha'$) in the effective action should trace back to a single extension of the generalised tangent bundle. A particular higher-derivative coupling may belong to different families and be derived from different extensions.

 \vskip 0.3cm
\noindent
The paper is organised as follows. After reviewing some facts on generalised geometry in the beginning of section \ref{sec:GG}, we show in subsection \ref{sec:GGBianchi} how its dimensional reductions give rise to a non-trivial Bianchi identity.
Section \ref{sec:nonAb}  is concerned with the incorporation of non-Abelian gauge groups in generalised geometry and the construction of the extended  generalised tangent bundle.  In particular, in subsection \ref{sec:nonAbelian} we discuss how a non-trivial Bianchi identity arises in generalised geometry, and discuss the symmetries in subsection \ref{sec:nonAbelian1}. In subsection \ref{sec:nonAbelian2} we then discuss an analogous construction for the gravitational part of the Bianchi identity and in subsection \ref{sec:redgauge} we discuss the importance of choice of connections for symmetry properties, such as T-duality. The significance of such a choice is further elaborated on in our discussion of the effective actions in section \ref{sec:gravconnection} where we consider the coupling of the $N=1$ ten-dimensional supergravity to Yang-Mills. We start by identifying the physical degrees of freedom and by constructing the generalised connections in subsection \ref{sec:localO}, which are then used to write down the action and equations of motion in subsection \ref{sec:sugraEQ}. Section \ref{sec:het-corr} is devoted to the $\alpha'$ corrections in heterotic strings.  In particular, in subsection  \ref{sec:het} we explain why $\Omega^-$ is the necessary (composite) connection for the gravitational part of the Bianchi identity. The equations of motion up to order $\sim \alpha'$ are discussed in subsection \ref{sec:hetEQ}, while the inclusion of higher order $\alpha'$ corrections in the effective action is the subject of subsection \ref{sec:higher}. We also show how the consistency of the extended generalised tangent bundle prevents the appearance of corrections linear in $\alpha'$ in type II theories. For the fluency of presentation, we have placed this discussion in Appendix~\ref{sec:typeII}. Section \ref{sec:qb!} discusses how the construction of the extended generalised tangent bundle reflects on corrections to  the Courant bracket.  In section \ref{sec:gen} we show how to construct a general formalism by which generalised geometry can be viewed as an infinite-dimensional version of the embedding tensor formalism. Finally we conclude with a speculative section \ref{sec:NS5}, where the modifications of the Bianchi identity due to the presence of NS5 sources are discussed.

\section{Basics of generalised geometry}
\label{sec:GG}

We first very briefly review some basic notions in generalised
geometry \cite{Hitchin,Gualtieri}, which can be thought of as giving
a geometric formulation of the NSNS sector of type II theories, namely
the metric $g$, B-field $B$ and dilaton $\phi$. The bosonic symmetries
of the theory are
\begin{equation*}
\begin{aligned}
   \text{1. } & \text{diffeomorphisms} \quad
      (g,B,\phi) \mapsto (g+{\cal L}_v g, B+{\cal L}_v B, \phi+
         {\cal L}_v\phi)\ ,\\
   \text{2. } & \text{gauge transformations}
      \quad (g,B,\phi) \mapsto (g,B - \dd\lambda,\phi) . \  \\
\end{aligned}
\end{equation*}
where ${\cal L}$ is the Lie derivative. The symmetries are therefore
generated by vectors $v$ and one-forms $\lambda$. These are combined
in the generalised tangent bundle into generalised vectors. More
precisely, the generalised tangent
bundle $E$ on a manifold $M$ is defined as a particular exact extension of
$TM$ by $T^*M$
\begin{equation}
\label{eq:Edef}
   0 \longrightarrow T^*M \longrightarrow E
      \stackrel{\pi}{\longrightarrow} TM \longrightarrow 0 .
\end{equation}
Generalised vectors are sections of $E$. Locally they can be written
as $V=v+\lambda$ or $V=(v,\lambda)$ where $v\in \Gs{TM}$ and $\lambda \in
\Gs{T^*M}$. In going from one coordinate patch $U_i$ to another $U_j$, we
have to first make the usual patching of vectors and one-forms, and
then give a further patching describing how $T^*M$ is fibered over
$TM$ in $E$. The latter is given by
\begin{equation}
\label{eq:EpatchGG}
\begin{aligned}
   v_{(i)} &= v_{(j)}\ , \\
   \lambda_{(i)} &= \lambda_{(j)}
      - \iota_{v_{(j)}}\dd\lambda_{(ij)} \ .
\end{aligned}
\end{equation}
This corresponds to the patching
\beq
\label{eq:Bpatch}
B_{(i)}=B_{(j)} - \dd \lambda_{(ij)} \ .
\eeq
There is a natural $O(n,n)$ pairing on $E$, given by
\begin{equation} \label{eta}
\langle V , W \rangle
    = \tfrac{1}{2}\iota_v \rho + \tfrac{1}{2}\iota_w \lambda \ ,
    \quad {\rm where} \ V=(v,\lambda)\ , \ W = (w, \rho)  \ .
\end{equation}

The differentiable structure on $E$ is encoded in the generalised Lie
derivative defined as
\begin{equation}\label{eq:genLie}
 \Lgen_{V} W = [v,w] + {\cal L}_v \rho - \iota_w \dd \lambda  \
\end{equation}
This has the Leibniz property
\begin{equation}\label{eq:relLieCourant}
   \comm{\Lgen_V}{\Lgen_W}
     = \Lgen_{\Lgen_V W} = \Lgen_{\Bgen{V}{W}} \ ,
\end{equation}
where the Courant bracket is defined as the antisymmetrization
\begin{equation}\label{eq:Courant}
 \Bgen{V}{W}
    = \tfrac{1}{2}\left( \Lgen_V W - \Lgen_W V \right)
    = [v,w] + {\cal L}_v \rho - {\cal L}_w \lambda
       - \tfrac12 \dd  (\iota_v \rho -\iota_w \lambda ) \ .
\end{equation}
The inner product (\ref{eta}) is invariant under the generalised Lie derivative, i.e.\
\begin{equation}\label{eq:scalarLie}
\langle \Lgen_{V} W , U \rangle + \langle W , \Lgen_{V}  U \rangle = {\cal L}_v \langle W , U \rangle \ .
\end{equation}
In addition to diffeomorphisms, the action $ \lambda \to \lambda -
\iota_v B$ is an automorphism of the generalised Lie derivative if
$\dd B = 0$. If however $H=\dd B \ne 0$, it transforms
as
\begin{equation} \label{eq:genLietrafo}
  \Lgen_{\e^B V} (\e^B W) = \e^B  ( \Lgen_{V} W + \iota_v \iota_w H) \ .
\end{equation}
Note that here $H=\dd B$, and correspondingly
\beq \label{BianchitypeII}
\dd H=0 \ .
\eeq

The NSNS fields define a generalised metric $G$. To include the
dilaton, one considers the weighted generalised tangent space
$\tilde{E}=\det T^*M\otimes E$ which admits a natural action of
$O(n,n)\times\bbR^+$, the $\bbR^+$ corresponding to a rescaling of the
weighting. A generalised metric $G$ is an $O(n)\times O(n)$
structure. It defines an isomorphism $\tilde{E}\simeq E$ together with
a splitting $E=C_+\oplus C_-$, with the orthonormal generalised
vielbeins on $C_\pm$ given by
\begin{equation}
\begin{aligned}
\label{eq:OdOdframe}
   \hat{E}^+_a &=
      \ee^{-2\phi}\sqrt{-g}\,\big(
        \hat{e}^+_a + e^+_a + i_{\hat{e}^+_a}B \big) , \\
   \hat{E}^-_{\bar{a}} &=
      \ee^{-2\phi}\sqrt{-g}\,\big(
         \hat{e}^-_{\bar{a}} -e^-_{\bar{a}} +
         i_{\hat{e}^-_{\bar{a}}}B \big) ,
\end{aligned}
\end{equation}
where $\hat{e}^{\pm}$ are orthonormal bases for $g$ in $TM$ and $e^{\pm}$
their duals. The two $O(n)$ groups act separately on $\hat{E}^+$ and
$\hat{E}^-$ and
\begin{align}
   \GM{\hat{E}^+_a}{\hat{E}^+_b} &=  \norm{\vol_G}^2\delta_{ab} , &
   \GM{\hat{E}^-_{\bar{a}}}{\hat{E}^-_{\bar{b}}}
         &=  -\norm{\vol_G}^2\delta_{\bar{a}\bar{b}} , &
   \GM{\hat{E}^+_a}{\hat{E}^-_{\bar{b}}} &= 0 ,
\end{align}
where $\norm{\vol_G}=\ee^{-2\phi}\sqrt{-g}$. In the special case where the
two frames are aligned, so $e^+_a=e^-_a=e_a$, it is conventional to
also define generalised vielbeins of the form
\beq \label{genviel}
 \hat{E}_{A} = \ee^{-2\phi}\sqrt{-g}\left( \begin{aligned}
 e^a \\   \hat e_a + \iota_{e_a}  B
 \end{aligned} \right) \ .
\end{equation}
such that
\begin{equation}
   \GM{\hat{E}_A}{\hat{E}_B}
      = \frac12 \norm{\vol_G}^2
      \begin{pmatrix} 0 & \id \\ \id & 0 \end{pmatrix}
\end{equation}

In the next section we will see how by dimensional reduction we obtain
a non-zero right-hand side to (\ref{BianchitypeII}). We will do the
simple case of a circle reduction, which is enough to understand the
idea, and can be somewhat easily generalised to tori reductions.

\subsection{Circle reductions and Bianchi identities}
\label{sec:GGBianchi}

From a dimensional reduction we can obtain a right-hand side to the
Bianchi identity of the form given by the contribution from (an
Abelian) gauge bundle in the heterotic theory. This will allow us to
get some intuition on the concepts introduced in the next section.

Consider an $(n+1)$-dimensional manifold $\tilde{M}$ that is a circle
fibration with base $M$. The generalised geometry on $\tilde{M}$ has
an $O(n+1,n+1)$ structure invariant under a $U(1)$ action generated by a
vector $\del/\del_{\varphi}$, where locally $\varphi$ parametrises the
circle fibration. If we require all the fields to be independent of
$\varphi$, the metric on such a background takes the form
\begin{equation}
\dd s^2 = \dd s^2_{n} + \e^\rho (\dd \varphi + C_1)^2 \ ,
\end{equation}
where
\beq
G_2 = \dd C_1
\eeq
gives the field strength of the fibration. Moreover, we can also
decompose the field strength $H$ of the $B$-field as
\begin{equation}\label{eq:Hdecomp}
H = H_3 + H_2 \wedge (\dd \varphi + C_1) \ ,
\end{equation}
where $\iota_{\del \varphi} H_3=0$.

If we started with a trivial Bianchi identity (\ref{BianchitypeII}) in
$n+1$ dimensions, we find now for the $n$-dimensional fields
\begin{equation}
\dd H_3 = - H_2 \wedge G_2 \ , \qquad \dd H_2 = 0 \ ,
\end{equation}
i.e.\ a non-trivial Bianchi identity. If we define
\begin{equation}\label{eq:redefpm}
 F^\pm = \tfrac12 (H_2 \pm G_2) \ ,
\end{equation}
we end up with the Bianchi identity
\begin{equation}
\label{eq:modBI}
\dd H_3 = F^- \wedge F^- - F^+ \wedge F^+  \ ,
\end{equation}
which can be viewed as an Abelian version of \eqref{eq:hetBianchi}.
If we reduce the two-form gauge field $B$ itself, using
\begin{equation}
B = \tilde B_2 + B_1 \wedge (\dd \varphi + C_1) \
\end{equation}
we find
\begin{equation}\label{Habelian}
H_3 = \dd \tilde B_2 - B_1 \wedge G_2 \ , \qquad H_2=\dd B_1 \ .
\end{equation}
Under the redefinition \eqref{eq:redefpm} the first equality becomes
\begin{equation} \label{H3}
H_3 = \dd B_2 + A^- \wedge \dd A^- - A^+ \wedge \dd A^+  \ ,
\end{equation}
where we defined
\beq
B_2 = \tilde B_2 + A^- \wedge A^+ \ , \qquad A^\pm = \tfrac12 (B_1 \pm C_1)
\eeq
and we have $F^\pm = \dd A^\pm$.

The generalised vielbein (\ref{genviel}) now reads
\begin{equation}
 E_{A} = \left( \begin{aligned}
 e^a \\ \e^\rho (\dd \varphi + C_1) \\ \hat e_a + \iota_{e_a} \tilde B_2  \\ \e^{-\rho} (\del_\varphi + B_1)
 \end{aligned} \right) \ .
\end{equation}
Note that T-duality along $\varphi$ just amounts to an exchange of
$G_2$ and $H_2$ (and of course an invertion $\rho\to - \rho$). In
other words it maps $A^\pm \to \pm A^\pm$, while $H_3$ and $B_2$ are
invariant.

Recall that the $B$-field must be patched as in~\eqref{eq:Bpatch}. On
triple overlaps one has the gerbe structure
\begin{equation}
   \Lambda_{(ij)} + \Lambda_{(jk)} + \Lambda_{(ki)}
      = \dd \Lambda_{ijk}
\end{equation}
with $g_{ijk}=\exp(4\pi\alpha'\ii\Lambda_{ijk})$ satisfying
$g_{jkl}g^{-1}_{ikl}g_{ijl}g_{ijk}^{-1}=1$ so that $H$ is
quantised. Assuming all the gauge transformation parameters are
independent of $\varphi$ it is easy to see that this implies that the
$B_1$ bundle is trivial, and hence the Pontryagin classes $[F^-\wedge
F^-]$ and $[F^+\wedge F^+]$ are equal in cohomology, as required by the form
of~\eqref{eq:modBI}.

Forgetting about the circle as a physical direction, we can view the
above construction as $n$-dimensional generalised geometry on $M$ with two
$U(1)$ gauge groups. The symmetries are now
\begin{equation*}
\begin{aligned}
   \text{1. } & \text{diffeomorphisms} \ ,\\
   \text{2. } & \text{one-form transformations}
      \quad B \mapsto  B - \dd\lambda \ ,\\
   \text{3. } & \text{gauge transformations}  \quad (A^+,A^-) \mapsto (A^+ + \dd\Lambda^+,A^- + \dd\Lambda^-) \ .
\end{aligned}
\end{equation*}
Geometrically, the bundle in consideration becomes $TM \oplus
\mathbb{R} \oplus \mathbb{R}\oplus T^*M$, the two $\bbR$ factors corresponding to
the adjoint representation of the $U(1)$ groups, and there is a
scalar product
\begin{equation}
\label{eq:modscalar}
\langle V, W \rangle = \tfrac12\iota_v \sigma + \tfrac12\iota_w \lambda +  \Lambda^+ \Sigma^+ -  \Lambda^- \Sigma^- \ ,
\end{equation}
where we defined $V= (v, \Lambda^+, \Lambda^-,\lambda)$ and $W = (w, \Sigma^+, \Sigma^-, \sigma)$.
Furthermore, the generalised vielbein is given by
\begin{equation}
 E_A^M = \left( \begin{aligned}
 e^a_m \\ f^+ + f^+ A^+_m \\ f^- + f^- A^-_m \\  \hat e^m_a + (\iota_{e_a} \tilde B_2)_m  
 \end{aligned} \right) \ .
\end{equation}
where $f^\pm$ are some non-vanishing functions giving a basis for the
two $\bbR$ bundles.

We find a generalised geometry whose Bianchi identity is sourced by the first Pontryagin class of two $U(1)$ bundles. Similarly, one can also discuss the Bianchi identity with only one $U(1)$ bundle, giving rise to $B_n$ generalised geometry on a bundle $TM  \oplus \mathbb{R}\oplus T^*M$ \cite{Hitchin:Bn,Bn,Rubio}. In section~\ref{sec:nonAbelian} we will discuss non-Abelian gauge groups in generalised geometry by generalizing the bundle construction employed here to the non-Abelian case.

Let us now consider the generalised Lie derivative \eqref{eq:genLie}. After the circle reduction (assuming that the generalised vectors do not depend on $\varphi$) we find that the generalised Lie derivative splits into
\begin{equation}\label{eq:genLieAbel}
 \Lgen_V W = {\cal L}_v w + {\cal L}_v \rho - \iota_w \dd \lambda + 2 \Sigma^+ \dd \Lambda^+ - 2 \Sigma^- \dd \Lambda^- + {\cal L}_v \Sigma^+ - {\cal L}_w \Lambda^+  + {\cal L}_v \Sigma^- - {\cal L}_w \Lambda^-  \ .
\end{equation}
Notably, the one-form transformation has a non-trivial contribution from gauge transformations.
Note that the scalar product \eqref{eq:modscalar} obeys the relation
\eqref{eq:scalarLie} with this generalised Lie derivative
\eqref{eq:genLieAbel}. The Courant bracket then gets modified to
\begin{equation}\begin{aligned}
 \Bgen{V}{W} =   & [v,w] + {\cal L}_v \rho - {\cal L}_w \lambda  - \tfrac12 \dd  (\iota_v \rho -\iota_w \lambda) +  \Sigma^+ \dd \Lambda^+ - \Lambda^+ \dd \Sigma^+  -  \Sigma^- \dd \Lambda^- +  \Lambda^- \dd \Sigma^-\\&
+  {\cal L}_v \Sigma^+ - {\cal L}_w \Lambda^+  + {\cal L}_v \Sigma^- - {\cal L}_w \Lambda^-  \ .
\end{aligned}\end{equation}

Before dimensional reduction, shifts by $B$ were automorphisms of the Courant bracket if $\dd B=0$. This together with shifts in $A_1$ such that $F_2= \dd A_1 = 0$ give the automorphisms of \eqref{eq:genLieAbel}, that are transformations induced by $B_2$ and $A^\pm$,
\begin{equation} \label{patchingabelian} \begin{aligned}
B_2\ :\ & (v,\lambda) \to (v,\lambda - \iota_v B_2) \ ,\\
& \Lambda^\pm \to \Lambda^\pm , \\
A^\pm \ : \ & (v,\lambda) \to (v,\lambda \pm 2 \Lambda^\pm A^\pm \mp (\iota_v A^\pm) A^\pm) \ , \\
& \Lambda^\pm \to \Lambda^\pm - \iota_v A^\pm \ .
\end{aligned}\end{equation}
From \eqref{eq:genLietrafo} we can also compute the transformation behavior if $B_2$ and $A$ are not closed, that is
\begin{equation} \label{twisting1} \begin{aligned}
 \Lgen_{\e^{B_2} V} (\e^{B_2} W) = & \e^{B_2} \left(\Lgen_{V} W + \iota_v \iota_w \dd B_2 \right) \ , \\
 \Lgen_{\e^{A^+}V} (\e^{A^+} W) = & \e^{A^+} \left( \Lgen_{V} W +  2 (\Sigma^+ \iota_v - \Lambda^+ \iota_w)F^+ -\iota_v \iota_w (A^+ \wedge \dd A^+) + \iota_v \iota_w (\dd A^+)\right) \ , \\
 \Lgen_{\e^{A^-}V} (\e^{A^-} W) = & \e^{A^-} \left( \Lgen_{V} W  - 2 (\Sigma^- \iota_v - \Lambda^- \iota_w)F^- +\iota_v \iota_w (A^- \wedge \dd A^-)+ \iota_v \iota_w (\dd A^-) \right) \ .
\end{aligned}\end{equation}
Puting everything together we can defined the twisted generalised Lie derivative
\begin{equation}
\hat \Lgen_{V} W = \Lgen_{V} W + 2 (\Sigma^+ \iota_v - \Lambda^+ \iota_w)F^+ - 2 (\Sigma^- \iota_v - \Lambda^- \iota_w)F^- + \iota_v \iota_w H_3 + \iota_v \iota_w F^+ + \iota_v \iota_w F^-)  \ .
\end{equation}
with the correction terms combining into the gauge-invariant field strengths $H_3$ and $F^\pm$.

The simple calculation presented here exemplifies many of the points we will see in the more general set-up.
We shall turn now to considering a non-Abelian gauge group $G$. Formally one may take $G$ to be a product group and have an $GL(n,\bbR)$ or $O(n)$ factor. Trying to identify this $GL(n,\bbR)$ factor with gravitational contributions to the Bianchi identity is associated to a number of complications, the discussion of which is the essential part of this paper.


\section{Non-Abelian gauge groups in generalised geometry}
\label{sec:nonAb}


In this section we will discuss different aspects of incorporating non-Abelian gauge symmetries in generalised geometry. This gives the geometrical basis for describing $N=1$ supergravity coupled to super Yang--Mills theory, and also $\alpha'$ corrections.

We have already seen how Abelian groups can appear from a reduction on a circle. For non-Abelian groups the basic construction is by reduction on a general group manifold. In the mathematics literature, this is known as a ``generalised reduction''~\cite{reduction}. The resulting extended generalised tangent space with the corresponding generalised Lie derivative is known as a ``transitive Courant algebroid''. That such algebroids can be constructed by reduction was first observed by Severa~\cite{severa} (see also~\cite{csx} for a discussion) and first discussed in the generalised geometric context in~\cite{Hitchin:Bn,Bn} and~\cite{Rubio}. It was specifically applied to the heterotic theory in~\cite{GF,Baraglia}. We note that a Courant algebroid is a vector bundle $E$ with a map $\pi:E\to TM$, a Dorfman bracket (or generalised Lie derivative) $\Lgen_VW\in \Gs{E}$ and an (indefinite) metric $\met{U}{V}$, with the properties
\begin{equation}
\label{eq:CA-def}
\begin{aligned}
   (1) && & \comm{\Lgen_U}{\Lgen_V}W = \Lgen_{\Lgen_UV}W , \\
   (2) && & \met{\Lgen_UV}{W}+\met{V}{\Lgen_UW}
       = \iota_{\pi(U)} \dd\met{V}{W} , \\
   (3) && & \Lgen_VV = 2 \dd \met{V}{V} .
\end{aligned}
\end{equation}
Using the metric one can define the dual map $\frac{1}{2}\pi^*:T^*M\to E$, and in the last equation this is used to interpret the one-form on the right-hand side as an element of $E$. The Courant algebroid is transitive if the map $\pi$ is surjective, and is exact if it is transitive and the sequence $T^*M\to E\to TM$ defined by $\pi$ and $\frac{1}{2}\pi^*$ is exact. It is crucially the third condition that means that transitive Courant algebroids encode a non-trivial Bianchi identity for $H$.

In the physics literature, the reduction picture was discussed in~\cite{dnpw,David}, the form of the generalised Lie derivative and the local generalised geometry in the DFT context is given in~\cite{DFT-het}, and in~\cite{GM} it is shown how such structures appear by effectively a generalised reduction.

\subsection{Symmetry algebra and patchings}
\label{sec:nonAbelian}

We saw in the previous section how a reduction on $S^1$ led to a theory with two $U(1)$ gauge fields, which could be described by an extension of conventional generalised geometry, with a generalised Lie derivative given by~\eqref{eq:genLieAbel}. Necessarily the total Pontryagin class of the bundle was zero. We now review how this extends to the non-Abelian case with group $G$. One way to view the structure is again as a reduction from higher dimensions, this time choosing a space which is a fibration with fibre $G$, that is a principle bundle.\footnote{The $S^1$ reduction gave $U(1)^2$ with a trivial $U(1)$ factor from the reduction of the $B$-field gerbe. In the following, we will simply keep a single $G$ factor.} However, one can also just construct the generalised geometry from first principles using the non-Abelian analogues of the objects defined in the previous section, or equivalently, the symmetries of $N=1$ supergravity coupled to non-Abelian gauge theory. For simplicity this is the approach we will use. Again we will see that the total Pontryagin class of the gauge bundle has to vanish. This is encoded in the Bianchi  identity of the form
\begin{equation} \label{eq:Bianchi_gauge}
\dd H_3 =  \tr F \wedge F\ ,
\end{equation}
where $F=\dd A + A \wedge A$  is now the field strength of some non-Abelian gauge field $A$. This will be the starting point of the construction.

The non-Abelian version of (\ref{H3}) realizing  \eqref{eq:Bianchi_gauge} is
\begin{equation}
   H_3 = \dd B_2 +  \omega_3(A) \ ,
\end{equation}
where the Chern--Simons three-form $\omega_3(A)$ is given by
\begin{equation}\label{eq:CS}
    \omega_3(A) = \tr\left( A\wedge F - \tfrac{1}{3}A^3 \right) \ .
\end{equation}
Under a finite gauge transformation
\beq \label{finitegaugetr}
A\mapsto A'=gAg^{-1}+  g\dd g^{-1} \ ,
\eeq
one finds
\begin{equation}
   \omega_3(A') = \omega_3(A) + \dd\tr\left(g^{-1}\dd g\wedge A\right)
      - \tfrac{1}{3}\tr\left(g\dd g^{-1}\right)^3 \ .
\end{equation}
The final term is closed,\footnote{Actually it must be since
$\dd\omega_3(A)=\tr F\wedge F$ is a gauge invariant four-form.}
thus locally we can write
\begin{equation}
   \tfrac{1}{3}\tr\left(g\dd g^{-1}\right)^3 = \dd \mu_2(g) \ ,
\end{equation}
and hence to keep $H$ invariant, one actually transforms
\begin{equation} \label{Bdef}
   B \mapsto B' = B -  \tr\left(g^{-1}\dd g\wedge A\right)
         + \mu_2(g)\ .
\end{equation}

Infinitesimally the three transformations are generated by vectors $v$
for diffeomorphisms, one-forms $\lambda$ and Lie algebra elements
$\Lambda$.

As before, we can combine these
into a single object $V=v+\Lambda+\lambda$ such that
\begin{equation}
\label{eq:variations}
\begin{aligned}
   \delta_V A_{(i)} &= \mathcal{L}_{v_{(i)}}A_{(i)}
      -  \dd\Lambda_{(i)} + \comm{\Lambda_{(i)}}{A_{(i)}}\ , \\
   \delta_V B_{(i)} &= \mathcal{L}_{v_{(i)}}B_{(i)}
      - \dd \lambda_{(i)}
      -   \tr\left(\dd\Lambda_{(i)}\wedge A_{(i)}\right)\ .
\end{aligned}
\end{equation}
Eqs. (\ref{finitegaugetr}) and (\ref{Bdef})  imply that the patching (\ref{eq:EpatchGG}) is now of the form (cf. (\ref{patchingabelian}) for the Abelian case, without diffeomorphisms)
\begin{equation}
\begin{aligned}
   v_{(i)} = & v_{(j)}\ , \\
   \dd\lambda_{(i)} = & \dd\lambda_{(j)}
      - \mathcal{L}_{v_{(j)}}\dd\lambda_{(ij)}
      +2 \dd\tr\left(\Lambda_{(j)}g^{-1}_{(ij)}\dd g_{(ij)}\right)
      +  \dd\tr\left[(\mathcal{L}_vg_{(ij)})\dd g_{(ij)}^{-1}\right]\\ & +  \dd (\iota_{v_{(j)}} \mu_2(g_{(ij)}) )\ , \\
   \Lambda_{(i)}  = & g_{(ij)}\Lambda_{(j)} g_{(ij)}^{-1}
      + g_{(ij)}\mathcal{L}_{v_{(j)}}g_{(ij)}^{-1}\ .
\end{aligned}
\end{equation}
There is an ambiguity in integrating the second relation but we can
\emph{choose} it such that
\begin{equation}
\label{eq:Epatch}
\begin{aligned}
   v_{(i)} &= v_{(j)}\ , \\
   \lambda_{(i)} &= \lambda_{(j)}
      - \iota_{v_{(j)}}\dd\lambda_{(ij)}
      +2 \tr\left(\Lambda_{(j)} g^{-1}_{(ij)}\dd g_{(ij)}\right)
      + \tr\left[(\mathcal{L}_vg_{(ij)})\dd g_{(ij)}^{-1}\right]+  \iota_{v_{(j)}} \mu_2(g_{(ij)})\ , \\
   \Lambda_{(i)} & = g_{(ij)}\Lambda_{(j)} g_{(ij)}^{-1}
      + g_{(ij)}\mathcal{L}_{v_{(j)}} g_{(ij)}^{-1} \ .
\end{aligned}
\end{equation}
The first two terms in the second line are the usual
patching one takes for the generalised tangent space, eq. (\ref{eq:EpatchGG}). The last three
terms are corrections which describe how the $B$-patching is twisted
by the gauge bundle.

We can view~\eqref{eq:Epatch} as defining a generalised heterotic
tangent space
\begin{equation}
   E \simeq  TM\oplus \adj P_G \oplus  T^*M
\end{equation}
where $\adj P_G$ is the adjoint bundle with fibres in the Lie algebra $\mathfrak{g}$ of $G$. The bundle is naturally defined as an extension using the patchings~\eqref{eq:Epatch}. Crucially, we have seen that such an extension is only possible if the gauge bundle has trivial first Pontryagin class.

This space has the natural metric, invariant across patches, given by the non-abelian generalisation of~\eqref{eq:modscalar} 
\begin{equation}
\label{eq:metricgauge}
   \met{V}{W} = \tfrac12\iota_v \sigma+\tfrac12\iota_w \lambda 
      + \tr \Lambda \Sigma \ .
\end{equation}
We note that this gives $E$ the structure of a $O(n,n+\dim{G})$ bundle, where $\dim{G}$ is the dimension of $G$. The differential structure is given by the generalised Lie derivative (\ref{eq:genLieAbel}) where the last term gets an extra contribution with repect to the non-Abelian version, i.e. we have
\begin{equation} \label{eq:genLiegauge}
   \Lgen_V W = {\cal L}_v w + {\cal L}_v \Sigma - {\cal L}_w \Lambda  +
       \comm{\Lambda}{\Sigma} + {\cal L}_v \rho - \iota_w \dd \lambda + 2 \tr \Sigma \dd \Lambda  \ ,
\end{equation}
where  $V=v+\Lambda+\lambda$. Similarly, the Courant bracket is given by the antisymmetrisation $\Bgen{V}{W} = \tfrac12 \Lgen_V W - \tfrac12 \Lgen_W V$. It takes the form
\begin{equation}
\label{eq:Bgengauge}
 \Bgen{V}{W} =    [v,w] +  {\cal L}_v \Sigma - {\cal L}_w \Lambda +  \comm{\Lambda}{\Sigma} +
 {\cal L}_v \rho - {\cal L}_w \lambda  - \tfrac12 \dd  (\iota_v \rho -\iota_w \lambda) +  \tr (\Sigma \dd \Lambda - \Lambda \dd \Sigma)  \ .
\end{equation}
and since $ \comm{\Lgen_U}{\Lgen_V} = \Lgen_{\Bgen{U}{V}}$ it reproduces the algebra of the
variations~\eqref{eq:variations} by
$\comm{\delta_V}{\delta_{W}}=\delta_{\Bgen{V}{W}}$. It is easy to check that $\Lgen_VW$ satisfies the conditions~\eqref{eq:CA-def} and so defines a transitive Courant algebroid.


\subsection{Splittings and automorphisms}
\label{sec:nonAbelian1}

Given the symmetries of the underlying supergravity we expect that the construction is invariant under under combinations of diffeomorphisms, closed $B$-shifts, pure-gauge ``$A$-shifts'' and global gauge transformations
\begin{equation} \label{eq:automorphisms}
\begin{aligned}
   \text{$B$-shifts:}& \quad \lambda \mapsto \lambda - \iota_v B &&
      \qquad \dd B=0 \ , \\
   \text{$A$-shifts:}& \quad
      \begin{cases}
         \lambda &\mapsto \lambda + 2  \tr(\Lambda A) -  \tr[(\iota_v A)A] \ ,  \\
         \Lambda &\mapsto \Lambda - \iota_vA \ ,
      \end{cases}
       && \qquad \dd A + A\wedge A =0 , \\
   \text{global gauge:}& \quad
       \Lambda \mapsto g \Lambda g^{-1}
       && \qquad \dd g = 0 \ .
\end{aligned}
\end{equation}
These preserve the metric~\eqref{eq:metricgauge} and furthermore are automorphisms of the generalised Lie derivative~\eqref{eq:genLiegauge}. 

More generally, we can view generic choice of $A$ and $B$ as providing a splitting of the generalised tangent space, that is, an explicit isomorphism $E\simeq TM\oplus \adj{P_G}\oplus T^*M$, which allows us to take a combination of vectors, gauge parameters and one-forms and lift them into a section of $E$. Writing this map as $\ee^B\ee^A$ we find, for generic $A$ and $B$ shifts, the generalised Lie derivative transforms as
\begin{equation} \label{eq:genLiegauge_shifts}
   \Lgen_{\e^B \e^A V} \e^B \e^A V' = \e^B \e^A \hat \Lgen_{V} V' \ ,
\end{equation}
where
\begin{equation}\begin{aligned}\label{eq:genLietwist}
  \hat \Lgen_{V} V' = & \comm{v}{v'} + \mathcal{L}_v \lambda'
      - \iota_{v'} \dd\lambda +2  \tr (\Lambda' D \Lambda) + 2  \tr (\Lambda' \iota_v F) - 2  \tr (\Lambda \iota_{v'} F)  \\ &
      + \iota_v \iota_{v'} H  + \comm{\Lambda}{\Lambda'} + \iota_v D \Lambda' - \iota_{v'}D\Lambda + \iota_v \iota_w F \ ,
\end{aligned} \end{equation}
and where we introduced the gauge covariant derivative $D \Lambda = \dd \Lambda +[A,\Lambda]$ and the field strength $F = \dd A + A\wedge A$.

Note also, that there are vector duals of $A$-shifts, which we might call
$\alpha$-shifts
\begin{equation}
   \text{$\alpha$-shifts:} \quad
       \begin{cases}
       v \mapsto v + 2 \tr(\Lambda\alpha) - \tr[(\iota_\alpha
           \lambda)\alpha]\ , \\
       \Lambda \mapsto \Lambda - \iota_\alpha \lambda \ ,
       \end{cases}
\end{equation}
which are the analogues of $\beta$-transformations (and similarly fail
to preserve the bracket).
These are the complementary elements in $O(n,n+\dim G)$.

\subsection{Extended generalised tangent space for  $G \times \tilde{G}$}
\label{sec:nonAbelian2}

As the first step towards understanding the gravitational contribution $ \tfrac{1}{4 } \alpha' \tr R\wedge R$  to the heterotic Bianchi identity \eqref{eq:hetBianchi} let us consider an extension by a product group  $G\times \tilde{G}$, where $\tilde{G}=GL(n,\bbR)$. We will also now include explicit factors of $\alpha'$. Of course by treating the gravitational connection $\omega = \tilde A$ as independent $\tilde G$ connection, we introduce unphysical degrees of freedom. For now, let us not worry about that and simply take advantage of the fact and in the ``new" Bianchi identity
\begin{equation}
\label{eq:fakeBI}
 \dd H = \tfrac14 \alpha' \left[\tr F \wedge F  - \tr {\tilde F}\wedge {\tilde F}\right] \ ,
\end{equation}
the gravitational contribution takes the familiar form and can be treated like a gauge one. In particular
\begin{equation}
   H = \dd B + \tfrac14 \alpha' (\omega_3( A) - \omega_3(\tilde A)) ,
\end{equation}
where $\omega_3$ is given by \eqref{eq:CS}.
The gauge fields $A$ and $\tilde A$ transform under gauge transformations as
\begin{equation}
A\mapsto gAg^{-1} + g\dd g^{-1} \ , \qquad \tilde A \mapsto \tilde g \tilde A \tilde g^{-1}+\tilde g\dd \tilde g^{-1} \ .
\end{equation}
This implies that $B$ transforms as
\begin{equation} \label{eq:gaugetrafoB}
 B \to B - \tfrac14 \alpha' \left(\tr\left(g^{-1}\dd g\wedge A\right) - \tr\left(\tilde g^{-1}\dd \tilde g\wedge \tilde A\right) \right)
         + \tfrac14 \alpha' \left(\mu_2( g) - \mu_2(\tilde g)\right) \ .
\end{equation}
For the patching of the generalised tangent bundle we find (similar to \eqref{eq:Epatch})
\begin{equation}\label{eq:Epatch2gauge}
\begin{aligned}
   v_{(i)} =& v_{(j)} \ , \\
   \lambda_{(i)} =& \lambda_{(j)}
      - \iota_{v_{(j)}}\dd\lambda_{(ij)}
      + \tfrac12 \alpha' \left( \tr\left(\Lambda_{(j)} g^{-1}_{(ij)}\dd g_{(ij)}\right) - \tr\left(\tilde \Lambda_{(j)} \tilde g^{-1}_{(ij)}\dd \tilde g_{(ij)}\right) \right) \\ &
      + \tfrac14 \alpha' \left( \tr\left((\mathcal{L}_vg_{(ij)})\dd g_{(ij)}^{-1}\right) - \tr\left((\mathcal{L}_v \tilde g_{(ij)})\dd \tilde g_{(ij)}^{-1}\right) \right) + \tfrac14 \alpha' \iota_{v_{(j)}} (\mu_2( g_{(ij)}) - \mu_2(\tilde g_{(ij)}) )\ , \\
   \tilde \Lambda_{(i)}  =& g_{(ij)}\Lambda_{(j)} g_{(ij)}^{-1}
      + g_{(ij)}\mathcal{L}_{v_{(j)}} g_{(ij)}^{-1} \ ,\\
   \Lambda_{(i)}  =& g_{(ij)}\Lambda_{(j)} g_{(ij)}^{-1}
      + g_{(ij)}\mathcal{L}_{v_{(j)}} g_{(ij)}^{-1}  \ .
\end{aligned}
\end{equation}
The scalar product reads
\begin{equation}
\label{eq:metric2gauge}
   \met{V}{V'} = \tfrac12\iota_v \lambda' + \tfrac12\iota_{v'} \lambda + \tfrac14 \alpha' (\tr ( \Lambda  \Lambda') - \tr (\tilde \Lambda \tilde \Lambda')) \ ,
\end{equation}
and is invariant under the patching given in \eqref{eq:Epatch2gauge}. The generalised Lie
derivative is given by
\begin{equation} \label{eq:genLie2gauge}\begin{aligned}
   \Lgen_V V' = & \comm{v}{v'} + \mathcal{L}_v \lambda'
      - \iota_{v'} \dd\lambda + \tfrac14 \alpha' (\tr  \Lambda' \dd \Lambda - \tr \tilde\Lambda' \dd\tilde\Lambda ) \\ &
      + \comm{\Lambda}{\Lambda'} + \iota_v \dd \Lambda' - \iota_{v'}\dd\Lambda + \comm{\tilde \Lambda}{\tilde \Lambda'} + \iota_v \dd \tilde \Lambda' - \iota_{v'}\dd\tilde \Lambda\ .
\end{aligned}\end{equation}
and the Courant bracket by
\begin{equation}
\label{eq:Bgen2gauge}
\begin{aligned}
   \Bgen{V}{V'} = & \comm{v}{v'}
       + \mathcal{L}_v\lambda' - \mathcal{L}_{v'}\lambda
       - \tfrac{1}{2}\dd\left(\iota_v\lambda' - \iota_{v'}\lambda\right)
       \\ &
       + \tfrac18 \alpha' (\tr ( \Lambda' \dd \Lambda) - \tr ( \Lambda \dd \Lambda') - \tr (\tilde\Lambda' \dd\tilde\Lambda) + \tr (\tilde\Lambda \dd\tilde\Lambda'))\\ &
       + \comm{\Lambda}{\Lambda'}
       + \left(\iota_{v}\dd\Lambda' - \iota_{v'}\dd\Lambda \right)
       + \comm{\tilde \Lambda}{\tilde \Lambda'} + \left( \iota_v \dd \tilde \Lambda' - \iota_{v'}\dd\tilde \Lambda \right) \ .
\end{aligned}
\end{equation}

The relative minus sign between the two terms in \eqref{eq:fakeBI} is a matter of trace conventions for $G$ and $\tilde G$ respectively and is reflected in our construction of $E$.  However, we shall see that once chosen this relative sign propagates to the the effective action and the $F^2$ and ${\tilde F}^2$ terms will also have opposite signs. This will be addressed in sections~\ref{sec:gravconnection} and~\ref{sec:het-corr}.

We shall now have a preliminary discussion of the effect of choosing $\tilde{G}=GL(n,\bbR)$ on the Courant brackets and how imposing extra symmetry properties, such as T-duality, covariance may restrict the so-far unconstrained connections.

\subsection{Local data: isometries and T-duality}
\label{sec:redgauge}

In section~\ref{sec:GGBianchi} we reviewed the circle reduction in generalised geometry. We will now redo the analysis in the extended case, and study the effect of the presence of a gauge group in the higher-dimensional theory. Once again, the Bianchi identity \eqref{eq:Bianchi_gauge} will be our starting point and we introduce a shorthand for the right-hand side
\begin{equation}\label{eq:Bianchi_general}
\dd H = X (F) \ .
\end{equation}
We have already seen how the generalised geometry restricts the topology of the gauge bundle. The reason for redoing this exercise is to study the role of the connections themselves, and see how they can be restricted by the constraint of realising of T-duality. This leads us in to the local constructions of sections~\ref{sec:gravconnection} and~\ref{sec:het-corr}.

As already explained,  $X(F)$ is exact, i.e.\
\begin{equation}
X(F) = \kappa \dd \omega(A) \ ,
\end{equation}
where $\omega(A)$ in the case of $X = \kappa \tr(F^2)$ is given by the Chern-Simons term \eqref{eq:CS}.
Using the decomposition \eqref{eq:Hdecomp} and
\begin{equation}
X(F)= X_4(F) + X_3(F) \wedge (\dd \varphi + C_1) \ ,
\end{equation}
the Bianchi identity \eqref{eq:Bianchi_general} translates into
\begin{equation}
\dd H_3 = X_4 - H_2 \wedge G_2 \ , \qquad \dd H_2 = X_3 \ .
\end{equation}
We can replace $X_4$ and $X_3$ by the components $\omega_3$ and $\omega_2$ of the Chern-Simons term $\omega(A)$, with
\begin{equation}\label{eq:dimdecomp_CS}
\omega(A)= \omega_3 + \omega_2 \wedge (\dd \varphi + C_1) \ .
\end{equation}
We find that
\begin{equation}\label{eq:Bianchi_red}
\dd H_3  = \kappa  \dd \omega_3 - \tilde G_2 \wedge G_2 \ ,
\end{equation}
and
\begin{equation}
\dd \tilde G_2 = 0 \ ,
\end{equation}
where we defined
\begin{equation}\label{eq:tildeG2}
\tilde G_2= H_2 - \kappa \omega_2 \ .
\end{equation}

In order to understand the properties of  $\omega_3$ and $\omega_2$, we dimensionally reduce the gauge connection $A$, i.e.\
\begin{equation}
A = A_1 + a (\dd \varphi + C_1) \ ,
\end{equation}
where $A_1$ is the lower-dimensional connection while the Wilson lines $a$ transform tensorially under lower-dimensional gauge transformations, i.e.\
\begin{equation}
A_1 \to g A_1 g^{-1} + g \dd g^{-1} \ , \qquad a \to g a g^{-1} \ ,
\end{equation}
where $g$ does not depend on $\varphi$.
From \eqref{eq:dimdecomp_CS} and \eqref{eq:CS} we find
\begin{equation}\begin{aligned}
\omega_3 = & \tr( A_1 \wedge \dd A_1 + \tfrac23 A_1^3) + \tr(a A_1) \wedge G_2 \ , \\
\omega_2 = & \tr(A_1 \wedge \dd a + a \dd A_1 + 2 a A_1 \wedge A_1) + \tr(a^2) G_2 \ .
\end{aligned}\end{equation}
By defining the gauge-covariant derivative $D$ on $a$ as
\begin{equation}
Da = \dd a + [A_1,a] \ ,
\end{equation}
we can rewrite $\omega_2$ in a tensorial way as
\begin{equation}
 \omega_2 = \tr(DDa) + \tr(a^2) G_2 \ ,
\end{equation}
which means $\omega_2$ is gauge invariant and therefore globally defined, which in turn means that $\tilde G_2$ in \eqref{eq:tildeG2} is globally defined. Moreover, defining
\begin{equation}
\tilde H_3 = H_3 - \tfrac12 \tr(Da) \wedge G_2 \ ,
\end{equation}
we can solve \eqref{eq:Bianchi_red} by
\begin{equation}
\tilde H_3 = \dd \tilde B_2 + \kappa \omega(A_1)  + \omega(A^-) - \omega(A^+) \ ,
\end{equation}
where $A^\pm$ are the gauge connections for $F^\pm = \tfrac12 (\tilde G_2 \pm G_2)$. Furthermore, the two-form gauge field $\tilde B_2$ appears in the decomposition of $B$ as
\begin{equation}
B = (\tilde B_2 + \tfrac12 \tr(a) G_2) + B_1 \wedge (\dd \varphi + C_1) \ .
\end{equation}
In total, the Bianchi identity of the lower-dimensional three-form field strength $\tilde H_3$ reads
\begin{equation}
\dd \tilde H_3 = \kappa \tr (F_2^2) + \tr ((G^-)^2) - \tr ((G^+)^2) \ .
\end{equation}

We can now consider T-duality on the circle. This will exchange $G_2$ and $\tilde  G_2$, and we see that the three-form $\tilde H_3$ has been constructed such that it remains invariant. Furthermore, also the gauge fields $A_1$ must transform such that $\tr (F_2^2)$ is T-duality invariant. Examples are $A_1$ itself being invariant or anti-invariant. The Wilson lines $a$ on the other hand can have a very non-trivial transformation behaviour.

If $A$ is simply a connection in $G$ these constraints are rather mild. However, let us now turn to the gravitational connection $\tilde A$ in $\tilde{G}=GL(n,\bbR)$. The invariance of the horizontal  part of  $\tr { \tilde F} \wedge {\tilde F}$ immediately implies  that $\tilde A$ cannot be the Levi-Civita connection but must involve $H$, since T-duality exchanges metric and $B$-field components. In fact as shown in \cite{LM} this is satisfied provided $\tilde F$ is computed using the torsionful connections
\begin{equation}
\Omega^\pm = \omega^{\tiny{\rm LC}} \pm \tfrac12 H \ ,
\end{equation}
where $\omega^{\tiny{\rm LC}}$ is the Levi-Civita connection. The torsion of $\Omega^\pm$ is just $\mp H$.
Note that the T-duality considerations do not fix the relative sign in $\Omega^\pm$ but do imply that the connection cannot be generic. As already mentioned, this is also the connection that is naturally chosen by world-sheet supersymmetry. The goal of sections~\ref{sec:gravconnection} and~\ref{sec:het-corr} is to show how the generalised geometry singles out the choice of $\Omega^\pm$.


\section{Action, connections and more}
\label{sec:gravconnection}

Thus far we have discussed how the extension of conventional
generalised geometry by a non-Abelian gauge group captures the global
topology that encodes the non-trivial Bianchi identity for $H$. To
connect to the actual supergravity theories, we need to construct the
local dynamics.

In this section we shall present an ``all-at-once'' construction of
effective actions based on the generalisations of the Lichnerowitz
theorem. This will reproduce the standard ten-dimensional, $N=1$
supergravity coupled to super Yang--Mills with gauge group
$G$~\cite{CM}, which has particle content
\begin{equation}
\begin{aligned}
   \text{gravity multiplet:} \quad &
      \left\{ g, B, \phi ; \psi_m, \rho \right\} \\
   \text{gauge multiplet:} \quad &
      \left\{ A ; \zeta \right\} .
\end{aligned}
\end{equation}
As usual in generalised geometry, the construction is to consider a
refined structure on $E$ and then define the appropriate compatible
generalised connection.

In the next section, we shall adapt this general construction to the
specific examples of lowest order $\alpha'$ corrections in heterotic
string and see the importance of the local data. A related
construction based on~\cite{CSW1} was given
in~\cite{GF}, see also~\cite{Baraglia}. However, when applied to the heterotic theory,
this required the $\Omega^-$ connection in the Bianchi
identity~\eqref{eq:hetBianchi} to be put in by hand. Using the
construction here, as we will see in the next section, $\Omega^-$
will appear naturally.

\subsection{Local structure group}
\label{sec:localO}

As in the previous section, we identify a generalised tangent space
$E\simeq TM\oplus \adj P_G \oplus T^*M$, for a generic gauge group
$G$ whose Lie algebra is $\mathfrak{g}$. Given an invariant metric on
$\mathfrak{g}$, denoted by the trace, there is a natural interior
product on $E$
\begin{equation}
   \GM{v+\Lambda+\lambda}{v'+\Lambda'+\lambda'}
      =\tfrac12\left(\iota_v \lambda' + \iota_{v'} \lambda\right)
         +\tfrac{\alpha'}{4}\tr\Lambda\Lambda' ,
\end{equation}
providing an isomorphism $E^*\simeq E$. For the weighted generalised
vector bundle $\tilde{E}=(\det T^*M)\otimes E$ one can then define a
corresponding bundle $\tilde{F}$ of frames, orthonormal up to a
conformal factor. By definition, this is an
$O(n+\dim G,n)\times\bbR^+$ principal bundle.

Given the fields $(g,B,A,\phi)$ one can define the set of
frames\footnote{Nearly all notation and conventions   follow~\cite{CSW1}.}
\begin{equation}
\begin{aligned}
\label{eq:splitframe}
   \hat{E}^+_a &= \ee^{-2\phi}\sqrt{-g}\,\big(\hat{e}^+_a + e^+_a + \iota_{\hat{e}^+_a}B +
      \iota_{\hat{e}^+_a}A - \tfrac{\alpha'}{4}\tr A\,\iota_{\hat{e}^+_a}A \big) ,\\
   \hat{E}^+_{\alpha} &= \ee^{-2\phi}\sqrt{-g}\,\big(\sqrt{\tfrac{4}{\alpha'}}t_{\alpha}
      -  \sqrt{\alpha'}\,\tr t_{\alpha}A\big) ,\\
   \hat{E}^-_{\bar{a}} &= \ee^{-2\phi}\sqrt{-g}\,\big(\hat{e}^-_{\bar{a}}
      -e^-_{\bar{a}} + \iota_{\hat{e}^-_{\bar{a}}}B
      + \iota_{\hat{e}^-_{\bar{a}}}A
      - \tfrac{\alpha'}{4} \tr A\,\iota_{\hat{e}^-_{\bar{a}}}A\big) ,
\end{aligned}
\end{equation}
where $\hat{e}^{\pm}$ are orthonormal bases in $TM$, $e^{\pm}$ their
dual, and
$[t_{\alpha},t_{\beta}]=f_{\alpha\beta}{}^{\gamma}t_{\gamma}$ give a
basis for $\mathfrak{g}$, with $\tr t_{\alpha}t_{\beta}=
\delta_{\alpha\beta}$. We have
\begin{equation}
\begin{aligned}
   \GM{\hat{E}^+_a}{\hat{E}^+_b} &= \norm{\vol_G}^2\delta_{ab} , \\
   \GM{\hat{E}^+_\alpha}{\hat{E}^-_{\beta}}
         &= \norm{\vol_G}^2\delta_{\alpha\beta} , \\
   \GM{\hat{E}^-_{\bar{a}}}{\hat{E}^-_{\bar{b}}}
         &=  -\norm{\vol_G}^2\delta_{\bar{a}\bar{b}} ,
\end{aligned}
\end{equation}
where $\norm{\vol_G}=\ee^{-2\phi}\sqrt{-g}$, and all other inner
products vanishing. Following~\cite{GF, Baraglia}, one can view these
frames as defining a generalised metric which splits $E=\tilde{C}_+\oplus
\tilde{C}_-$, and reduces the structure group to $O(n+\dim G)
\times O(n)$, with the respective frames\footnote{Note that given the
   $O(n+\dim G)$ frame is not generic: one could also take
   rotated combinations of $\hat{E}^+_a$ and $\hat{E}^+_{\alpha}$.}
\begin{equation}
   \hat{E}_A =
      \begin{cases}
         \left(\hat{E}^+_a,\hat{E}^+_{\alpha}\right) &
            O(n+\dim G)_+ \\
         \hat{E}^-_{\bar{a}} & O(n)_-
      \end{cases} .
\end{equation}
We actually want to reduce the structure group on $E$ further to
disentangle $G$ from the $O(n)$ inside $\tilde{C}_+$. In other words, we pick
a substructure $\tilde{P}\subset\tilde{F}$ of the generalised frame
bundle which is an $O(n)\times G \times O(n)$ principal bundle. This
further splits $E = C_+\oplus C_{\mathfrak{g}}\oplus C_-$  with the
respective frames
\begin{equation}
   \hat{E}_A =
      \begin{cases}
         \hat{E}^+_a & O(n)_+ \\
         \hat{E}^+_{\alpha} & G  \\
         \hat{E}^-_{\bar{a}} & O(n)_-
      \end{cases} .
\end{equation}
It is in this setting that we would like to present the
supergravity equations in terms of generalised connections, much in
the same way as for the type II case in~\cite{CSW1}.

In~\cite{CSW1} it was shown that all supergravity equations can be
written in terms of generalised connections which are compatible with
the local $O(n)\times O(n)$ structures and are torsion-free, and that
even though there are several such connections, the supergravity
equations are uniquely determined, as they should
be. In~\cite{GF} a similar approach was taken for the
heterotic theory, using the local $O(n+\dim G)\times O(n)$
structure discussed above. The supergravity equations could again be
formulated using torsion-free, compatible generalised connections,
however, the form of the $R$-curvature in the Bianchi
identity~\eqref{eq:hetBianchi} has to be put in by hand. A priori, it
is the curvature of an arbitrary connection on a generic $O(n)$
bundle. The condition that the bundle is that coming from $TM$ and
that the relevant connection is $\Omega^-$ as given
in~\eqref{eq:Omegaminus} does not come from the construction.

To overcome this problem we will instead consider generalised
connections that are compatible with the $O(n)\times G \times O(n)$
structure $\tilde{P}=\tilde{P}_+\oplus \tilde{P}_G\oplus
\tilde{P}_-$. The consequence is that we can no longer
require that generalised connections compatible with the reduced
structure be torsion-free,  as for a generic choice of $\tilde{P}$
(i.e. a generic field configuration) there  exists a non-vanishing
intrinsic torsion. To see this, let $D$ and $D'$ be two arbitrary
compatible generalised connections. Fix $D$ and define the
tensor $\Sigma = D' - D$. Then by varying $D'$ we have that $\Sigma$
spans
\begin{equation}
   K = E \otimes
       (\Lambda^2C_+\oplus \adj\tilde{P}_G\oplus\Lambda^2C_-),
\end{equation}
where $\adj\tilde{P}_G$ is the adjoint bundle with fibres
in the Lie algebra $\mathfrak{g}$ and we have used we used
$\adj\tilde{P}_\pm\simeq \Lambda^2 C_{\pm}$. Now the torsion of a
connection is defined by the bracket~\eqref{eq:genLiegauge} and is
an element~\cite{CSW1}
\begin{equation}
   T(D) \in \Gs{W}, \quad \text{where } W= \Lambda^3E\oplus E .
\end{equation}
Therefore we can define the associated map for
the difference of the torsions, $\tau: K \rightarrow W$,
such that
\begin{equation}
   \tau(\Sigma) = T(D') - T(D) .
\end{equation}
It is easy to check that the intrinsic torsion space
$W_{\text{int}}=\coker\tau$ is non-trivial,
\begin{equation}
   W_{\text{int}} = C_+\otimes C_{\mathfrak{g}}\otimes C_-.
\end{equation}
We thus have that, for a generic structure, every compatible
connection will have some torsion. Using the
frame~\eqref{eq:splitframe}, we can calculate the intrinsic torsion
explicitly and find it takes the form
\begin{equation}
\label{eq:int-tor}
   \Tint_{\bar{a}b\gamma}= 
      -\tfrac12 \sqrt{\alpha'}F_{\bar{a}b\gamma} .
\end{equation}
Thus we see that requiring a torsion-free connection sets $F=0$, which
is too strong a condition.

Instead of requiring the full torsion-free condition, we can take the
weaker physical conditions that the connections that appear in the
supergravity equations are uniquely determined. We consider first the
fermionic fields. Using the $O(n)\times G \times O(n)$ structure we
identify the fields\footnote{Note that we are using
   the redefined dilatino $\rho=\gamma^m\psi_m-\lambda$
   where $\lambda$ is the conventional dilatino field. Also, for the ten-dimensional theory one would need to correctly identify the chiralities of the fermion fields but in order to keep the discussion completely general we will not be explicit about this here.}
\begin{equation}\label{eq:FermionFields}
\begin{aligned}
   \text{gravitino:} & && \psi_a \in \Gs{C_+\otimes S(C_-)} , \\
   \text{gaugino:} & &&
      \zeta_\alpha \in \Gs{C_\mathfrak{g}\otimes S(C_-)} , \\
   \text{dilatino:} & && \rho \in \Gs{S(C_-)} ,
\end{aligned}
\end{equation}
where $S(C_-)$ is the spin-bundle for the $\Spin(n)$ group on $C_-$
(we assume throughout that our manifold is spin). The supersymmetry
parameter $\epsilon$ is similarly a section of $S(C_-)$. Comparing
with the structure of type II theories discussed in~\cite{CSW1},
we require that the following operators are uniquely determined, since
they will appear in either the supersymmetry transformations or the
fermion equations of motion
\begin{equation}
\begin{aligned}
\label{eq:SugraOps}
   \gamma^{\bar{a}}D_{\bar{a}}\rho, \quad &
      D^a\psi_a, &
      D^{\alpha}\zeta_{\alpha},\quad & \in \Gs{S(C_-)},\\
   \phantom{\gamma^{\bar{a}}D_{\bar{a}}\rho,\quad} & D_a \rho, &
      \gamma^{\bar{a}}D_{\bar{a}}\psi_a, \quad
      & \in \Gs{C_+\otimes S(C_-)},\\
   \phantom{\gamma^{\bar{a}}D_{\bar{a}}\rho,\quad} & D_{\alpha}\rho, &
      \gamma^{\bar{a}}D_{\bar{a}}\zeta_{\alpha}, \quad
      & \in \Gs{C_{\mathfrak{g}}\otimes S(C_-)}.
\end{aligned}
\end{equation}
Solving for the compatible connection we find that this implies
that the torsion of $T(D)$ is restricted to lie in the subspace
\begin{equation}
   T(D) \in  \Gs{W_{\text{restr}}},
\end{equation}   
where
\begin{equation*}
W_{\text{restr}} = \Lambda^3C_+ \oplus \Lambda^3C_{\mathfrak{g}}
       \oplus \big(C_+\otimes \adj \tilde{P}_G \big)
       \oplus \big(C_{\mathfrak{g}}\otimes\Lambda^2C_+\big)
       \oplus \big(C_+\otimes C_{\mathfrak{g}}\otimes C_- \big)\subset \Lambda^3 E \oplus E,
\end{equation*}
with the last term being the intrinsic torsion, which is independent of the
choice of $D$. Using the frames~\eqref{eq:splitframe}, we can solve
explicitly for $D$. If $\nabla=\partial+\omega+A$ is the covariant derivative associated with the combined
Levi--Civita and gauge connection, then acting on a generalised vector $W =
w_+^a\hat{E}^+_a + w_{\mathfrak{g}}^{\alpha}\hat{E}^+_{\alpha} +
w_-^{\bar{a}}\hat{E}^-_{\bar{a}} \in\Gs{E}$,  we have
\begin{subequations}
\label{eq:OdddLC}
\begin{align}
    D_a w_+^b
       &= \nabla_a w_+^b
           - \tfrac{2}{n-1}\big(
              \delta_a{}^b \partial_c\phi
              -\eta_{ac}\partial^b\phi \big)w_+^c
           + Q_a{}^b{}_c w_+^c  ,\label{eq:OdddLC1} \\
    D_{\alpha}  w_+^b
       &= Q_{\alpha}{}^b{}_a w_+^{a},\label{eq:OdddLC2} \\
    D_{\bar{a}} w_+^b
       &= \nabla_{\bar{a}} w_+^b
          - \tfrac{1}{2}H_{\bar{a}}{}^b{}_cw_+^c,
          \label{eq:OdddLC3}\\
    D_a  w_{\mathfrak{g}}^{\beta}
       &= \nabla_a w_{\mathfrak{g}}^{\beta}
          + Q_a{}^{\beta}{}_{\gamma}w_{\mathfrak{g}}^{\gamma},
          \label{eq:OdddLC4}\\
    D_{\alpha}  w_{\mathfrak{g}}^{\beta}
       &= Q_{\alpha}{}^{\beta}{}_{\gamma}w_{\mathfrak{g}}^{\gamma},
       \label{eq:OdddLC5}\\
    D_{\bar{a}}  w_{\mathfrak{g}}^{\beta}
       &= \nabla_{\bar{a}} w_{\mathfrak{g}}^{\beta},
       \label{eq:OdddLC6}\\
    D_a w_-^{\bar{b}}
       &= \nabla_a w_-^{\bar{b}}
           + \tfrac{1}{2}H_a{}^{\bar{b}}{}_{\bar{c}}w_-^{\bar{c}} ,
           \label{eq:OdddLC7}\\
    D_{\alpha} w_-^{\bar{b}}
       &= -\tfrac{1}{2}\sqrt{\alpha'}
          F^{\bar{b}}{}_{\bar{a}}{}_{\alpha}w_-^{\bar{a}} ,
          \label{eq:OdddLC8}\\
    D_{\bar{a}} w_-^{\bar{b}}
       &= \nabla_{\bar{a}} w_-^{\bar{b}}
           + \tfrac{1}{6}H_{\bar{a}}{}^{\bar{b}}{}_{\bar{c}}w_-^{\bar{c}}
           - \tfrac{2}{n-1}\big(
              \delta_{\bar{a}}{}^{\bar{b}} \partial_{\bar{c}}\phi
              - \eta_{\bar{a}\bar{c}}\partial^{\bar{b}}\phi
              \big)w_-^{\bar{c}}
           + Q_{\bar{a}}{}^{\bar{b}}{}_{\bar{c}} w_-^{\bar{c}} ,
           \label{eq:OdddLC9}
\end{align}
\end{subequations}
for any $Q\in\Gs{E\otimes\adj\tilde{P}}$ satisfying
\begin{equation}
\begin{aligned}
   Q_{a\bar{b}\bar{c}}&=0, & &&
   Q_{\alpha\bar{b}\bar{c}}&=0, & &&
   Q_{\bar{a}bc}&=0, & &&
   Q_{\bar{a}\beta\gamma}&=0, \\
   Q_{[\bar{a}\bar{b}\bar{c}]}&=0, & &&
   Q_{\bar{a}}{}^{\bar{a}}{}_{\bar{b}}&= 0, & &&
   Q_{\alpha}{}^{\alpha}{}_{\beta}&=0, & &&
   Q_{a}{}^{a}{}_b&= 0,
\end{aligned}
\end{equation}
and otherwise arbitrary. This means that $Q$ will in general
contribute to the torsion of $D$, for instance the fully antisymmetric
component $Q_{[abc]}$ will be in the torsion if it is non-vanishing.
In addition, again note that irrespective of the particular choice of
$Q$, all connections $D$ will have an intrinsic torsion given by
$\Tint_{\bar{a}b\gamma}= -\tfrac12 \sqrt{\alpha'}F_{\bar{a}b\gamma}$,
simply due to the fact that they are compatible with the reduced
structure $\tilde{P}$. Crucially, however, the supergravity
operators~\eqref{eq:SugraOps} are independent of the particular choice
of $D$, since all $Q$ contributions will drop out
of~\eqref{eq:SugraOps}.\footnote{As an alternative way of stating this,
   while we have been thinking of the unconstrained $Q$ as an element
   of $U\subset K = E\otimes \text{ad} \tilde{P}$, we could instead
   consider it as a map $\mathcal{Q}: S\oplus J\oplus J_{\mathfrak{g}}
   \rightarrow E\otimes (S\oplus J\oplus J_{\mathfrak{g}})$ where
   $S=S(C_-)$, $J=C_+\otimes S(C_-)$ and $J_{\mathfrak{g}}=
   C_{\mathfrak{g}}\otimes S(C_-)$.
   The linear projections in~\eqref{eq:SugraOps} on the other hand are
   a map $\mathcal{P}: E\otimes (S\oplus J\oplus J_{\mathfrak{g}})
   \rightarrow S\oplus J\oplus J_{\mathfrak{g}}$. The map
   $\mathcal{Q}$ is then such that precisely $\text{ker}\mathcal{P} =
   \text{Im} \mathcal{Q}$.

   The space $U$ spanned by $Q$ is thus set by the kernel of the
   projection to $S\oplus J\oplus J_{\mathfrak{g}}$. For the type II
   case, this space coincided with the kernel of the torsion map
   within the metric connections space. In this case, since
   the fermionic representations are smaller, the kernel of the
   projection is larger and includes some of the torsion.}

\subsection{Supergravity equations}
\label{sec:sugraEQ}

We next see how the generalised connection $D$ encodes the
equations of $N=1$ ten-dimensional supergravity coupled to super
Yang-Mills~\cite{CM}.

We start with the bosonic action. For any spinor
$\epsilon\in\Gs{S(C_-)}$, we can consider the following Bismut-type
equation:
\begin{equation}
\label{eq:newBismut}
   \gamma^{\bar{a}}D_{\bar{a}}\gamma^{\bar{b}}D_{\bar{b}}\epsilon
      = D^aD_a\epsilon
          + D^{\alpha}D_{\alpha}\epsilon
          - \tfrac14 S^-\epsilon ,
\end{equation}
Generically one might expect the right-hand side to contain scalar,
two-form and four-form terms of the form $T_{\bar{a}_a\dots
   \bar{a}_{2n}}\gamma^{\bar{a}_a\dots \bar{a}_{2n}}$ (since the
spinors are chiral the six-form terms can be written as
four-forms). However, remarkably, provided the Bianchi
identity~\eqref{eq:Bianchi_gauge} holds, only the scalar term $S^-$
survives. This scalar gives the dilaton equation of motion, or equivalently, the
bosonic action
\begin{equation}
\label{eq:BosonicAction}
   S^- = s +4 \nabla^2 \phi -4 (\partial \phi)^2
       - \tfrac{1}{12} H^2 -\tfrac{\alpha'}{8}\tr F^2 ,
\end{equation}
where $s$ is the Ricci scalar for the metric $g$.

The supersymmetry variations of the fermion fields are given by
\begin{equation}
\label{eq:FermVar}
\begin{aligned}
   \delta\psi_a &= D_a \epsilon
      = \nabla_a\epsilon+ \tfrac{1}{8}H_{a\bar{b}\bar{c}}
          \gamma^{\bar{b}\bar{c}} \epsilon , \\
   \delta\zeta_{\alpha} &= D_{\alpha}\epsilon
      = -\tfrac18 \sqrt{\alpha'}F_{\bar{a}\bar{b}\alpha}
          \gamma^{\bar{a}\bar{b}}\epsilon , \\
   \delta\rho &= \gamma^{\bar{a}}D_{\bar{a}}\epsilon
      = \gamma^{\bar{a}}\nabla_{\bar{a}} \epsilon
           + \tfrac{1}{24}H_{\bar{a}\bar{b}\bar{c}}
               \gamma^{\bar{a}\bar{b}\bar{c}}\epsilon
           - (\partial_{\bar{a}}\phi) \gamma^{\bar{a}}\epsilon ,
\end{aligned}
\end{equation}
while the fermionic action has the form
\begin{equation}
\begin{aligned}
\label{eq:FermionicAction}
   S_F = -\frac{1}{2\kappa^2}\int 2 \vol_G &\Big[
         \bar\psi^{a} \gamma^{\bar{b}} D_{\bar{b}} \psi_{a}
         + \bar\psi^{a} \gamma^{\bar{b}}
            \Tint_{\bar{b}}{}^{\alpha}{}_a \zeta_{\alpha}
         + 2 \bar\rho D_a \psi^{a}
         - \bar\rho \gamma^{\bar{a}} D_{\bar{a}} \rho  \\&
         + \bar\zeta^{\alpha} \gamma^{\bar{b}}
            D_{\bar{b}}\zeta_{\alpha}
         + \bar\zeta^{\alpha} \gamma^{\bar{b}}
            \Tint_{\bar{b}}{}^{a}{}_{\alpha}  \psi_{a}
         + 2 \bar\rho D_{\alpha} \zeta^{\alpha}
         \Big] ,
\end{aligned}
\end{equation}
where $\Tint$ is the intrinsic torsion~\eqref{eq:int-tor} which is
fixed by the choice of local structure $\tilde{P}$ (and independent of
choice of compatible $D$).

The Bismut-type formula~\eqref{eq:newBismut} can then be understood as
a consequence of the dilatino equations of motion closing into the
dilaton equation of motion under susy. The bosonic equations of motion
for $g,B,A$ will similarly contain torsion terms, with the
generalised Ricci given by
\begin{equation}
\begin{aligned}
\label{eq:Bosoniceoms2}
  \tfrac12 R_{\bar{a}b}\gamma^{\bar{a}}\epsilon &= [\gamma^{\bar{a}}D_{\bar{a}}, D_b]\epsilon+\gamma^{\bar{a}}\Tint_{\bar{a}}{}^{\beta}{}_b D_{\beta} \epsilon ,\\
  \tfrac12 R_{\bar{a}\beta}\gamma^{\bar{a}}\epsilon &= [\gamma^{\bar{a}}D_{\bar{a}}, D_{\beta}]\epsilon+\gamma^{\bar{a}}\Tint_{\bar{a}}{}^{b}{}_{\beta} D_{b} \epsilon ,
\end{aligned}
\end{equation}
and the $g$ and $B$ equations and $A$ equation corresponding to the
generalised Ricci flat conditions
\begin{equation}
   R_{\bar{a}a} = 0  \qquad \text{and} \qquad  R_{\bar{a}\alpha} = 0
\end{equation}
respectively. For the bosonic supersymmetry transformations we have
\begin{equation}
\label{eq:Evar}
\begin{aligned}
  \delta\hat{E}^+_a
      &= \bar{\epsilon} \rho \hat{E}^+_a
         - \bar{\epsilon} \gamma_{\bar{a}} \psi_a  \hat{E}^{-\bar{a}}, \\
\delta\hat{E}^+_{\alpha}
      &= \bar{\epsilon} \rho \hat{E}^+_{\alpha}
         - \bar{\epsilon} \gamma_{\bar{a}} \zeta_{\alpha}  \hat{E}^{-\bar{a}} , \\
  \delta\hat{E}^-_{\bar{a}}
      &= \bar{\epsilon}\rho \hat{E}^-_{\bar{a}}
         - \bar{\epsilon} \gamma_{\bar{a}} \psi_a  \hat{E}^{+a}
         - \bar{\epsilon} \gamma_{\bar{a}} \zeta_{\alpha}  \hat{E}^{+\alpha}  ,
\end{aligned}
\end{equation}
Note that all these equations are independent of $Q$ as required.

\section{$\alpha'$ corrections for heterotic strings}
\label{sec:het-corr}

We shall now turn to our principal example to see how the framework
introduced in the previous section can be used to describe the $\alpha'$
corrections in heterotic strings. A key point is that we will see the
Bianchi identity~\eqref{eq:hetBianchi} appears naturally with the
correct $\Omega^-$ connection. The same formalism can be used to
analyse type II theory and in particular to show why linear $\alpha'$
corrections are ruled out in this case. However we put this discussion
into an appendix~\ref{sec:typeII} for sake of fluency of the presentation.

\subsection{Structures and connections}
\label{sec:het}

The first step is to consider the construction of the previous section
with a product gauge group $G_1\times G_2$. We will eventually
identify $G_1$ with the gauge group $G$ and $G_2$ with $O(n)$ or $GL(n,\bbR)$ as in
section~\ref{sec:nonAbelian2}, but, for the moment, we will keep them
general. We will write their respective adjoint
indices as $\alpha = (\alpha^1,\alpha^2)$, with corresponding
generators $\{t_{\alpha^i}\}$. For the metric on the sum of the Lie
algebras we take the indefinite form $\tr = \tr_1 - \tr_2$ where
$\tr_i t_{\alpha^i}t_{\beta^i}= \eta^i_{\alpha^i\beta^i}$.

The generalised frame defining an $O(n)_+\times G_1\times G_2\times
O(n)_-$ structure and corresponding connection follow from the
expressions in the previous section. The generalised tangent space
decomposes as
\begin{equation}
   E = C_+ \oplus C_{\mathfrak{g}_1} \oplus C_{\mathfrak{g}_2} \oplus C_- ,
\end{equation}
We find that~\eqref{eq:splitframe} becomes
\begin{equation}
\begin{aligned}
\label{eq:splitframe2}
   \hat{E}^+_a &= \ee^{-2\phi}\sqrt{-g}\,\Big(\hat{e}^+_a
       + e^+_a + \iota_{\hat{e}^+_a}B + \iota_{\hat{e}^+_a}A^1
       + \iota_{\hat{e}^+_a}A^2
       \\ & \qquad \qquad \qquad \qquad \qquad
       - \tfrac{1}{4}\alpha'\tr_1 A^1\,\iota_{\hat{e}^+_a}A^1
       + \tfrac{1}{4}\alpha'\tr_2 A^2\,\iota_{\hat{e}^+_a}A^2 \Big) , \\
   \hat{E}^1_{\alpha^1} &= \ee^{-2\phi}\sqrt{-g}\,\Big(
       \sqrt{\tfrac{4}{\alpha'}}t_{\alpha^1}
       - \sqrt{\alpha'}\,\tr_1 t_{\alpha^1}A^1 \Big) , \\
   \hat{E}^2_{\alpha^2} &= \ee^{-2\phi}\sqrt{-g}\,\Big(
       \sqrt{\tfrac{4}{\alpha'}}t_{\alpha^2}
       + \sqrt{\alpha'}\,\tr_2 t_{\alpha^2}A^2 \Big) , \\
   \hat{E}^-_{\bar{a}} &= \ee^{-2\phi}\sqrt{-g}\,\Big(
       \hat{e}^-_{\bar{a}} - e^-_{\bar{a}}
       + \iota_{\hat{e}^-_{\bar{a}}}B + \iota_{\hat{e}^-_{\bar{a}}}A^1
       + \iota_{\hat{e}^-_{\bar{a}}}A^2
       \\ & \qquad \qquad \qquad \qquad \qquad
       -\tfrac{1}{4}\alpha' \tr_1 A^1\,\iota_{\hat{e}^-_{\bar{a}}}A^1
       + \tfrac{1}{4}\alpha' \tr_2 A^2\,\iota_{\hat{e}^-_{\bar{a}}}A^2
       \Big) ,
\end{aligned}
\end{equation}
where $A^i$ is a connection for $G_i$. The generalised connection becomes
\begin{subequations}\label{eq:OdggdLC}
\begin{align}
    D_a w_+^b
       &= \nabla_a w_+^b
           - \tfrac{2}{n-1}\big(
              \delta_a{}^b \partial_c\phi-\eta_{ac}\partial^b\phi \big)w_+^c
           + Q_a{}^b{}_c w_+^c  ,\label{eq:OdddLC12} \\
    D_{\alpha^i}  w_+^b &=
       Q_{\alpha^i}{}^b{}_a w_+^{a},\label{eq:OdddLC22} \\
    D_{\bar{a}} w_+^b &= \nabla_{\bar{a}} w_+^b -
       \tfrac{1}{2}H_{\bar{a}}{}^b{}_cw_+^c,
       \label{eq:OdddLC32}\\
    D_a  w_{\mathfrak{g}_i}^{\beta^i}
       &= \nabla^i_a w_{\mathfrak{g}_i}^{\beta^i}
       +Q_a{}^{\beta^i}{}_{\gamma^i}w_{\mathfrak{g}_i}^{\gamma^i},
       \label{eq:OdddLC42}\\
    D_{\alpha^i}  w_{\mathfrak{g}_j}^{\beta^j} &=
      Q_{\alpha^i}{}^{\beta^j}{}_{\gamma^j}w_{\mathfrak{g}_j}^{\gamma^j},
      \label{eq:OdddLC52}\\
    D_{\bar{a}} w_{\mathfrak{g}_i}^{\beta^i} &=
      \nabla^i_{\bar{a}} w_{\mathfrak{g}_i}^{\beta^i},
      \label{eq:OdddLC62}\\
    D_a w_-^{\bar{b}}
       &= \nabla_a w_-^{\bar{b}}
       + \tfrac{1}{2}H_a{}^{\bar{b}}{}_{\bar{c}}w_-^{\bar{c}} ,
       \label{eq:OdddLC72}\\
    D_{\alpha^i} w_-^{\bar{b}} &= \begin{cases}
       -\tfrac{1}{2}\sqrt{\alpha'}F^{1\bar{b}}{}_{\bar{a}}{}_{\alpha^1}
           w_-^{\bar{a}} & \text{if $i=1$} \\
       \tfrac{1}{2}\sqrt{\alpha'}F^{2\bar{b}}{}_{\bar{a}}{}_{\alpha^2}
           w_-^{\bar{a}} & \text{if $i=2$}
           \end{cases} , \label{eq:OdddLC82}\\
    D_{\bar{a}} w_-^{\bar{b}} &= \nabla_{\bar{a}} w_-^{\bar{b}}
       + \tfrac{1}{6}H_{\bar{a}}{}^{\bar{b}}{}_{\bar{c}}w_-^{\bar{c}}
       - \tfrac{2}{n-1}\big(
          \delta_{\bar{a}}{}^{\bar{b}} \partial_{\bar{c}}\phi
          - \eta_{\bar{a}\bar{c}}\partial^{\bar{b}}\phi \big)w_-^{\bar{c}}
       + Q_{\bar{a}}{}^{\bar{b}}{}_{\bar{c}} w_-^{\bar{c}} ,\label{eq:OdddLC92}
\end{align}
\end{subequations}

In order to match to the $\alpha'$ theory, we need to make a further
reduction of the structure group. The crucial step is that we identify the $G_2$ structure group with the $O(n)_+$ structure group. In the case were $G_2=O(n)$ we can just directly identity the two. If $G_2 = GL(n,\bbR)$, one can think of first reducing to $O(n)$ and then identifying with the $O(n)_+$ structure group. In this latter case, the reduction $O(n)\subset GL(n,\bbR)$ specifies a ``metric'', which is to be identified with the gravitational metric $g$. In either case we end up with a refined structure
\begin{equation}
   O(n)_+ \times G_1 \times G_2 \times O(n)_-
      \supset O(n)_+ \times G \times O(n)_- ,
\end{equation}
where now we identify $G_1=G$, the heterotic gauge group. Equivalently we identify
$C_{\mathfrak{g}_2}\simeq \Lambda^2C_+$, that is we consider
\begin{equation}
   E = C_+ \oplus C_{\mathfrak{g}} \oplus \Lambda^2C_+ \oplus C_- ,
\end{equation}
Concretely the $\alpha^2$ index now corresponds to an antisymmetric
pair of vector indices $[ab]$, and writing just $\alpha$ for the
$G_1=G$ index we
have
\begin{equation}
   \alpha^1 = \alpha , \qquad
   \alpha^2 = [ab] ,
\end{equation}
so, writing $w_{\mathfrak{g}_1}^{\alpha^1}=w^\alpha_{\mathfrak{g}}$ and
$w^{\alpha^2}_{\mathfrak{g}_2}=w_+^{ab}$, the components of the
generalised vector are $W=(w_+^a,w_{\mathfrak{g}}^\alpha,w_+^{ab},w_-^{\bar{a}})$.
We can then rewrite~\eqref{eq:OdddLC32} and~\eqref{eq:OdddLC62} as
\begin{equation}
\label{eq:matchD}
\begin{aligned}
    D_{\bar{a}} w_+^b
       &= \partial_{\bar{a}} w_+^b + \Omega^-_{\bar{a}}{}^b{}_cw_+^c, \\
   D_{\bar{a}}w_+^{ab}
       &= \partial_{\bar{a}}w_+^{ab}
           + A^2_{\bar{a}}{}^{a}{}_{c}w_+^{cb}
           + A^2_{\bar{a}}{}^{b}{}_{c}w_+^{ac} ,
\end{aligned}
\end{equation}
where $\Omega^-=\omega^{\text{LC}}-\frac{1}{2}H$, with
$\omega^{\text{LC}}$ the Levi--Civita connection.

We then make the further natural requirement that the generalised
connection is compatible with the $O(n)_+\times G\times O(n)_-$
structure. This requires that two derivatives in~\eqref{eq:matchD}
agree, that is
\begin{equation}
   A^2 = \Omega^- = \omega^{\text{LC}} - \tfrac{1}{2}H .
\end{equation}
With this identification the generalised derivative is given by
\begin{equation}
\label{eq:hetD}
\begin{aligned}
    D_a w_+^b
       &= \nabla_a w_+^b  - \tfrac{2}{n-1}\big(
              \delta_a{}^b \partial_c\phi-\eta_{ac}\partial^b\phi \big)w_+^c
           + Q_a{}^b{}_c w_+^c  ,\\
    D_{\alpha}  w_+^b &= Q_{\alpha}{}^b{}_a w_+^{a}, \\
    D_{aa'}  w_+^b &= Q_{aa'}{}^b{}_c w_+^c, \\
    D_{\bar{a}} w_+^b &= \nabla_{\bar{a}} w_+^b -
       \tfrac{1}{2}H_{\bar{a}}{}^b{}_cw_+^c, \\
    D_a  w_{\mathfrak{g}}^{\beta}
       &= \nabla_a w_{\mathfrak{g}}^{\beta}
       +Q_a{}^{\beta}{}_{\gamma}w_{\mathfrak{g}}^{\gamma}, \\
    D_{\alpha}  w_{\mathfrak{g}}^{\beta} &=
      Q_{\alpha}{}^{\beta}{}_{\gamma}w_{\mathfrak{g}}^{\gamma}, \\
    D_{aa'}  w_{\mathfrak{g}}^{\beta} &=
      Q_{aa'}{}^{\beta}{}_{\gamma}w_{\mathfrak{g}}^{\gamma}, \\
    D_{\bar{a}} w_{\mathfrak{g}}^{\beta} &=
      \nabla_{\bar{a}} w_{\mathfrak{g}}^{\beta}, \\
    D_a w_-^{\bar{b}}
       &= \nabla_a w_-^{\bar{b}}
       + \tfrac{1}{2}H_a{}^{\bar{b}}{}_{\bar{c}}w_-^{\bar{c}} , \\
    D_{\alpha} w_-^{\bar{b}} &=
       -\tfrac{1}{2}\sqrt{\alpha'}F^{\bar{b}}{}_{\bar{c}}{}_{\alpha}
           w_-^{\bar{c}} , \\
    D_{aa'} w_-^{\bar{b}} &=
       \tfrac{1}{2}\sqrt{\alpha'}R(\Omega^-)^{\bar{b}}{}_{\bar{c}aa'}
           w_-^{\bar{c}}  \\
    D_{\bar{a}} w_-^{\bar{b}} &= \nabla_{\bar{a}} w_-^{\bar{b}}
       + \tfrac{1}{6}H_{\bar{a}}{}^{\bar{b}}{}_{\bar{c}}w_-^{\bar{c}}
       - \tfrac{2}{n-1}\big(
          \delta_{\bar{a}}{}^{\bar{b}} \partial_{\bar{c}}\phi
          - \eta_{\bar{a}\bar{c}}\partial^{\bar{b}}\phi \big)w_-^{\bar{c}}
       + Q_{\bar{a}}{}^{\bar{b}}{}_{\bar{c}} w_-^{\bar{c}} ,
\end{aligned}
\end{equation}
where $R(\Omega^-)$ is the curvature of the $\Omega^-$
connection. Compatibility means that the $D_Mw_+^{bb'}$ expressions
follow directly from those for $D_Mw^b$, namely
\begin{equation}
\begin{aligned}
    D_a w_+^{bb'}
       &= \nabla_a w_+^{bb'} - \tfrac{2}{n-1}\big(
              \delta_a{}^b \partial_c\phi-\eta_{ac}\partial^b\phi \big)w_+^{cb'}
           + Q_a{}^b{}_c w_+^{cb'} \\ & \qquad
           - \tfrac{2}{n-1}\big(
              \delta_a{}^{b'} \partial_c\phi-\eta_{ac}\partial^{b'}\phi \big)w_+^{bc}
           + Q_a{}^{b'}{}_c w_+^{bc}  ,\\
    D_{\alpha}  w_+^{bb'} &= Q_{\alpha}{}^b{}_c w_+^{cb'}
        + Q_{\alpha}{}^{b'}{}_c w_+^{bc}, \\
    D_{aa'}  w_+^{bb'} &= Q_{aa'}{}^b{}_c w_+^{cb'}
       + Q_{aa'}{}^{b'}{}_c w_+^{bc} , \\
    D_{\bar{a}} w_+^{bb'} &= \nabla_{\bar{a}} w_+^{bb'}
       - \tfrac{1}{2}H_{\bar{a}}{}^b{}_cw_+^{cb'}
       - \tfrac{1}{2}H_{\bar{a}}{}^{b'}{}_cw_+^{bc} .
\end{aligned}
\end{equation}
As before the undetermined parts of that connection satisfy
\begin{equation}
\begin{aligned}
   Q_{a}{}^{a}{}_b&= 0 , & &&
   Q_{\bar{a}}{}^{\bar{a}}{}_{\bar{b}}&= 0, & &&
   Q_{\alpha}{}^{\alpha}{}_{\beta}&=0, & &&
   Q_{ac}{}^c{}_b &=0 .
\end{aligned}
\end{equation}

\subsection{Supergravity equations}
\label{sec:hetEQ}

As before the supergravity equations of motion and supersymmetry
variations all follow from the generalised connection $D$ defined
in~\eqref{eq:hetD}. The action comes from the Bismut-type
equation~\eqref{eq:newBismut}. Specialising to the case in hand of a
product group $G_1\times G_2=G\times O(n)$ with $\tr=\tr_1-\tr_2$, gives
\begin{equation}
\label{eq:newBismut2}
   \gamma^{\bar{a}}D_{\bar{a}}\gamma^{\bar{b}}D_{\bar{b}}\epsilon
       = D^aD_a\epsilon+D^{\alpha}D_{\alpha}\epsilon
           - D^{ab}D_{ab}\epsilon
           - \tfrac14 S^-\epsilon ,
\end{equation}
where we have fixed the normalisation of
the $O(n)$ trace $\tr t_{aa'}t_{bb'}=\delta_{ab}\delta_{a'b'}$. As
before, this equation only defines a scalar $S^-$ if the corresponding
Bianchi condition holds. Here this reads
\begin{equation}
\label{eq:hetBianchi*}
   \dd H = \tfrac{1}{4} \alpha' \left[\tr F\wedge F  - 
      \tr R(\Omega^-)\wedge R(\Omega^-)\right] ,
\end{equation}
reproducing the $\alpha'$ corrected condition with the correct
connection $\Omega^-$. Calculating the scalar $S^-$, one finds the
$\alpha'$ corrected bosonic action (or equivalently the dilaton
equation of motion)
\begin{equation}
\label{eq:BosonicAction22}
   S^- = s +4 \nabla^2 \phi -4 (\partial \phi)^2
	- \tfrac{1}{12} H^2 -\tfrac{1}{8}\alpha'\tr F^2
        + \tfrac{1}{8}\alpha'\tr R(\Omega^-)^2 ,
\end{equation}
with again the appropriate curvature-squared term.

The supersymmetry variation of the fermion fields
follow as before from the set of uniquely determined operators
\begin{equation}
\label{eq:FermVar2}
\begin{aligned}
   \delta\psi_a &= D_a \epsilon
      = \nabla_a\epsilon + \tfrac{1}{8}H_{a\bar{b}\bar{c}}
           \gamma^{\bar{b}\bar{c}} \epsilon , \\
   \delta\zeta_{\alpha} &= D_{\alpha}\epsilon
      = -\tfrac18 \sqrt{\alpha'}F_{\bar{a}\bar{b}\alpha}
           \gamma^{\bar{a}\bar{b}}\epsilon , \\
   \delta\rho &= \gamma^{\bar{a}}D_{\bar{a}}\epsilon
     = \gamma^{\bar{a}}\nabla_{\bar{a}} \epsilon
        + \tfrac{1}{24}H_{\bar{a}\bar{b}\bar{c}}
          \gamma^{\bar{a}\bar{b}\bar{c}}\epsilon
        -\gamma^{\bar{a}}(\partial_{\bar{a}}\phi)\epsilon .
\end{aligned}
\end{equation}
The formalism naturally has, in addition, a ``gaugino''
$\psi_{ab}\in\Gs{\Lambda^2C_+\otimes S(C_-)}$ for the $G_2=O(n)_+$
gauge group. The corresponding variation is
\begin{equation}
   \delta\psi_{ab} = D_{ab}\epsilon
      = \tfrac18
      \sqrt{\alpha'}R(\Omega^-)_{\bar{a}\bar{b}ab}
         \gamma^{\bar{a}\bar{b}}\epsilon .
\end{equation}
The obvious interpretation is that $\psi_{ab}$ is the standard
composite ``gravitino curvature''~\cite{BdR}, given to leading order
in the fermions by
\begin{equation}\label{eq:gravitino-curv-def}
   \psi_{mn} = \tfrac12 \sqrt{\alpha'}\left(\der_m \psi_n - \der_n \psi_m
      + \tfrac{1}{4} \Omega^+_{mpq} \gamma^{pq} \psi_{n}
      - \tfrac{1}{4} \Omega^+_{npq} \gamma^{pq} \psi_{m} \right).
\end{equation}
Calculating the corresponding variation one finds~\cite{BdR}
\begin{equation}
\label{eq:gravitino-curv}
   \delta\psi_{ab} = \tfrac18\sqrt{\alpha'}
      R(\Omega^+)_{ab\bar{a}\bar{b}}
         \gamma^{\bar{a}\bar{b}}\epsilon
      =D_{ab}\epsilon + \mathcal{O}(\alpha^{\prime }) .
\end{equation}
as required, where we have used the fact that
\begin{equation}
   R_{mnpq}(\Omega^-) = R_{pqmn}(\Omega^+) - \tfrac12 \dd H_{mnpq}
      = R_{pqmn}(\Omega^+) + \mathcal{O}(\alpha') .
\end{equation}
We will discuss briefly the question of the higher order $\alpha'$
corrections to~\eqref{eq:gravitino-curv} in the next section.

For the bosonic variations we have
\begin{equation}
\label{eq:Evar2}
\begin{aligned}
  \delta\hat{E}^+_a
      &= \bar{\epsilon} \rho \hat{E}^+_a
         - \bar{\epsilon} \gamma_{\bar{a}} \psi_a  \hat{E}^{-\bar{a}}, \\
  \delta\hat{E}_{\alpha}
      &= \bar{\epsilon} \rho \hat{E}^1_{\alpha}
         - \bar{\epsilon} \gamma_{\bar{a}} \zeta_{\alpha} \hat{E}^{-\bar{a}} , \\
  \delta\hat{E}^-_{\bar{a}}
      &= \bar{\epsilon}\rho \hat{E}^-_{\bar{a}}
         - \bar{\epsilon} \gamma_{\bar{a}} \psi_a  \hat{E}^{+a}
         - \bar{\epsilon} \gamma_{\bar{a}} \zeta_{\alpha}  \hat{E}^{\alpha}
         + \bar{\epsilon} \gamma_{\bar{a}} \psi_{ab}  \hat{E}^{ab}  ,
\end{aligned}
\end{equation}
and for the $O(n)$ basis
\begin{equation}
   \delta\hat{E}_{ab}
      = \bar{\epsilon} \rho \hat{E}_{ab}
         - \bar{\epsilon} \gamma_{\bar{a}} \psi_{ab}  \hat{E}^{-\bar{a}} ,
\end{equation}
which is equivalent to variation of the composite object
\begin{equation}
\label{eq:Omega-var}
  \tfrac12 \sqrt{\alpha'}\,\delta \Omega^-_{mab}
      = -\bar{\epsilon} \gamma_{m} \psi_{ab} + \mathcal{O}(\alpha^{\prime}),
\end{equation}
which indeed follows from the other variations. We are reproducing the
standard result~\cite{BdR} that $(\psi_{mn}, \Omega^-_m)$ indeed
transform as a gauge multiplet.

Finally we can also write the fermionic action including higher
derivative terms for $\psi_m$ following~\eqref{eq:FermionicAction}, as
\begin{equation}
\begin{aligned}
\label{eq:FermionicAction2}
   S_F = -\frac{1}{2\kappa^2}\int 2 \vol_G &\Big[
         \bar\psi^{a} \gamma^{\bar{b}} D_{\bar{b}} \psi_{a}
         + 2 \bar\rho D_a \psi^{a}
         - \bar\rho \gamma^{\bar{a}} D_{\bar{a}} \rho
         + 2 \bar\rho D_{\alpha} \zeta^{\alpha}
         - 2 \bar\rho D_{ab} \psi^{ab}  \\
         &+ \bar\zeta^{\alpha} \gamma^{\bar{b}} D_{\bar{b}}
            \zeta^1_{\alpha}
         - \bar\psi^{ab} \gamma^{\bar{b}} D_{\bar{b}} \psi_{ab} \\
         &+ \bar\zeta^{\alpha} \gamma^{\bar{b}}
            \Tint_{\bar{b}}{}^{a}{}_{\alpha} \psi_{a}
         - \bar\psi^{ab}\gamma^{\bar{b}}
            \Tint_{\bar{b}}{}^{c}{}_{ab}  \psi_{c}
         +\bar\psi^{a} \gamma^{\bar{b}}
            \Tint_{\bar{b}}{}^{\alpha}{}_a \zeta_{\alpha}
         -\bar\psi^{a} \gamma^{\bar{b}}
            \Tint_{\bar{b}}{}^{bc}{}_a \psi_{bc}
         \Big] ,
\end{aligned}
\end{equation}
where the intrinsic torsion terms are given by
\begin{equation}
   \Tint_{\bar{a}b\gamma}
      = -\tfrac{1}{2}\sqrt{\alpha'} F_{\bar{a}b\gamma} , \qquad
   \Tint_{\bar{a}bcc'}
      = \tfrac{1}{2}\sqrt{\alpha'} R(\Omega^-)_{\bar{a}bcc'} .
\end{equation}
The bosonic equations of motion then come from varying the
corresponding fermionic equations and are given by vanishing of
the components of the corresponding generalised Ricci tensor, defined
by
\begin{equation}
\begin{aligned}
\label{eq:Bosoniceoms3}
  \tfrac12 R_{\bar{a}b}\gamma^{\bar{a}}\epsilon
      &= [\gamma^{\bar{a}}D_{\bar{a}}, D_b]\epsilon
         + \gamma^{\bar{a}}\Tint_{\bar{a}}{}^{\beta}{}_b D_{\beta}
            \epsilon
         - \gamma^{\bar{a}}\Tint_{\bar{a}}{}^{cc'}{}_b D_{cc'}
            \epsilon , \\
   \tfrac12 R_{\bar{a}\beta}\gamma^{\bar{a}}\epsilon
      &= [\gamma^{\bar{a}}D_{\bar{a}}, D_{\beta}]\epsilon
         + \gamma^{\bar{a}}\Tint_{\bar{a}}{}^{b}{}_{\beta} D_{b}
            \epsilon , \\
   \tfrac12 R_{\bar{a},bb'}\gamma^{\bar{a}}\epsilon
      &= [\gamma^{\bar{a}}D_{\bar{a}}, D_{bb'}]\epsilon
         - \gamma^{\bar{a}}\Tint_{\bar{a}}{}^{c}{}_{bb'} D_{c}
            \epsilon .
\end{aligned}
\end{equation}
The full set of independent bosonic field equations are then
\begin{equation}
   S^- = R_{\bar{a}b} = R_{\bar{a}\beta} = 0 ,
\end{equation}
while the composite equation $R_{\bar{a},bb'}=0$ follows identically from the previous
to zeroth order in $\alpha'$, a well-known result from~\cite{BdR}.

\subsection{Higher orders in $\alpha'$ }
\label{sec:higher}

In~\cite{BdR}  a quartic action with $(\alpha')^3$ corrections was derived, and in fact  an iterative procedure that can in principle be used to generate an entire family of corrections  with arbitrarily high powers of $\alpha'$ was pointed out.  The procedure is based on the supersymmetric completion of the heterotic Bianchi identity \eqref{eq:hetBianchi}, and the key observation is that an identification of the auxiliary $O(n)$ Yang-Mills gauge fields with the gravitational degrees of freedom is possible.\footnote{Note that only the corrections to the Bianchi identity considered in \cite{BdR} are linear in $\alpha'$. The Bianchi identity itself does not change and is used  in the process of iteration.}  However, in order to make such an identification one has to equate the ``gaugino''  with the gravitino curvature~\eqref{eq:gravitino-curv-def}. On the other hand, as we have already seen, this identification is not quite consistent with supersymmetry variation we had for the gaugino  so far -- in fact it only works to order $\alpha'$
\begin{equation}
   \delta\psi_{ab} = D_{ab}\epsilon + \mathcal{O}(\alpha^{\prime }) .
\end{equation}
This is a consequence of the heterotic Bianchi identity \eqref{eq:hetBianchi}, since in order to make the identification of the gaugino one has to use the relation $R_{mnpq}(\Omega^-) -  R_{pqmn}(\Omega^+) = -12 \dd H_{mnpq}$. Explicitly, the correction to the variation of the gravitino curvature is then
\begin{equation}\label{eq:gravitino-curv-def2}
   \delta\psi_{ab} = D_{ab}\epsilon +\tfrac18\sqrt{\alpha'} \left(  \tfrac18\alpha'[\tr F\wedge F-\tr R(\Omega^-)\wedge R(\Omega^-)]_{ab\bar{a}\bar{b}}\right) \gamma^{\bar{a}\bar{b}}\epsilon.
\end{equation}

With this change, though, supersymmetry no longer closes on the action. Indeed, in~\cite{BdR} it was found that in order to restore supersymmetry it is necessary to introduce corrections to the supersymmetry transformations of the ``fundamental'' (i.e. non-composite) fields at order $(\alpha')^2$, and add new quartic terms to the action at order $(\alpha')^3$. The resulting theory is then supersymmetric, but only up to order $(\alpha')^3$ since the corrections to the gravitino variation mean that~\eqref{eq:gravitino-curv-def2} would get new $(\alpha')^3$ corrections -- which would then break supersymmetry again and require new higher-order corrections to the variations and the action, and so on. Clearly this is a process that can, in theory, be repeated to generate ever higher $\alpha'$ corrections.

In our context these issues can be viewed in a slightly different manner. Let us denote the corrected variation $\delta\psi_{ab} = D_{ab}\epsilon + \mathcal{O}(\alpha^{\prime }) = \hat{D}_{ab}\epsilon$. Our guiding principle for writing bosonic actions is the Bismut formula
\begin{equation}
   \gamma^{\bar{a}}D_{\bar{a}}\gamma^{\bar{b}}D_{\bar{b}}\epsilon
       - D^aD_a\epsilon+D^{\alpha}D_{\alpha}\epsilon
           + D^{ab}D_{ab}\epsilon
           =- \tfrac14 S^-\epsilon.
\end{equation}
However, if we were to simply replace $D_{ab}$ with $\hat{D}_{ab}$, the equation will no longer define just a ($(\alpha')^3$-corrected) scalar $S^-$, but it would also generate on the right-hand side a two-form and a four-form at order $(\alpha')^2$. (This is just an equivalent, and somewhat more straightforward way of seeing that the action is no longer supersymmetric.) We would then be naturally led to also modify the supersymmetry rules~\eqref{eq:FermVar2} at order $(\alpha')^2$ and introduce corrected operators $\hat{D}_a\epsilon$ and $\gamma^{\bar{a}}\hat{D}_{\bar{a}}\epsilon$ such that they precisely cancel the two-forms and four-forms in the Bismut formula, at least to order $(\alpha')^2$. As in~\cite{BdR}, this procedure would give rise to an iterative rule for generating higher $\alpha'$ corrections. We hope to provide some results in this direction in upcoming work.

We would like to emphasise that these all-orders-in-$\alpha'$ corrections to the effective action originate from a single extension of the generalised tangent bundle, that captures the contribution of the heterotic Bianchi identity \eqref{eq:hetBianchi*} linear in $\alpha'$. This aspect of our construction is fully concordant with the construction of \cite{BdR}.

\section{$\alpha'$ corrections and brackets}
\label{sec:qb!}

Our starting point for deriving the possible $\alpha'$ gravitation corrections has been to view $O(n)$ or $\GL(n,\bbR)$ as a ``gauge group'', build an appropriate extension of the generalised tangent bundle, and hence enlarge the set of symmetries parametrised by the generalised tangent space. The additional symmetries are simply local changes of frame. 

However, one could also consider a different parametrisation, where the additional frame rotation symmetries are induced by diffeomorphisms, and so are not independent parameters. This is particularly interesting when flows and the related Lie algebroids are considered. 

Vector fields generate diffeomorphisms by generating flows on manifolds. These diffeomorphisms in turn induce a $\GL(n,\bbR)$ action on the frame bundle of the manifold $F$. To see this, choose a frame $e^a$ and its dual $\hat e_b$. Under the flow generated by some vector field $v$ the frame $e^a$ will be rotated by an element in $\GL(n,\bbR)$, which to leading order is just given by its Lie derivative along $v$. Therefore, if we call the generator of this rotation $Lv$, we find for its components
\begin{equation}\label{eq:rotmatrix_explicit}
   (Lv)^a{}_b = \iota_{\hat e_b} {\cal L}_v e^a \ ,
\end{equation}
in terms of a basis of $\GL(n,\bbR)$ generators $\hat{u}^{b}{}_a$ which are given in terms of the frame $\hat{e}_a$ by $\hat{u}^b{}_a =e^b\otimes \hat{e}_a$. 
Note that \eqref{eq:rotmatrix_explicit} depends on the choice of a particular frame $e^a$. However, changing to a different frame corresponds to a global $\GL(n,\bbR)$ transformation which just transforms the generators $\hat{u}^{b}{}_a$ into each other. Therefore, the choice of frame is part of the gauge choice, and the construction will not depend on it. 

The idea now is to take the gravitational gauge bundle $\tilde{P}$ to have gauge group $\tilde{G}=GL(n,\bbR)$ and identify it with the frame bundle. We can then restrict the corresponding gauge parameter in the generalised tangent space to being of the form $\tilde \Lambda = Lv$.

Let us then discuss how this can be accomplished in more detail. For simplicity we will ignore for the moment the gauge group $G$, which we already know how to handle, and keep the focus on just the frame bundle. Generic infinitesimal $\tilde{G}$-equivariant automorphisms of $F$ can be understood in terms of the bundle $\mathcal{A}=TF/\tilde{G}\simeq TM\oplus\adj{F}$ (that is, the Atiyah algebroid for $F$) which inherits a Lie algebroid structure from the Lie bracket algebra on $TF$ itself. Elements of $\mathcal{A}$ are simply combinations of vectors $v$ and gauge parameters $\tilde{\Lambda}$ patched just as in~\eqref{eq:Epatch}. In fact, the extended generalised vector space $E$ is often thought of as an extension $T^*M\to E\to \mathcal{A}$.

Given a vector $v\in \Gs{TM}$ there is then a natural lift to an element $\tilde{v}\in\Gs{\mathcal{A}}$ given locally by
\begin{equation}
   \tilde{v} = v + Lv ,
\end{equation}
where the components of the gauge part $(Lv)^a{}_b$ are precisely the ones given in \eqref{eq:rotmatrix_explicit}. This has the property that $\widetilde{[v,w]}=[\tilde{v},\tilde{w}]$, where $[\tilde{v},\tilde{w}]$ is the bracket on the Lie algebroid $\mathcal{A}$. One can check that the gauge part $\tilde\Lambda=Lv$ indeed patches correctly. 
Note that one cannot think of $\tilde{v}$ as being a section of some sub-bundle of $\mathcal{A}$, since its definition involves derivatives of $v$ itself. It is merely an element of the subspace of sections of $\mathcal{A}$ defined to consist of those $(v, \tilde \Lambda)$ for which the condition $\tilde \Lambda = Lv$ with \eqref{eq:rotmatrix_explicit} holds.

As always, one needs a splitting to identify the isomorphism $\mathcal{A}\simeq TM\oplus\adj{F}$, which here is given by an arbitrary choice of $\GL(n,\bbR)$-connection $\omega$ for the frame with the associated covariant derivative $\nabla$. Using the identity $\iota_{\hat e_b} {\cal L}_v e^a = (\nabla_b v^a) - \iota_v\omega^a{}_b$ we write
\begin{equation}
   \tilde{v} = (v - \iota_v\omega) + (\nabla_b v^a)\hat{u}^b{}_a ,
\end{equation}
and identify $\nabla_b v^a$ as the element of $\adj{F}$. This is just an example of the gauge shifts we observed in \eqref{eq:automorphisms}.

One might wonder whether the same reparametrisation can be achieved for a choice $\tilde{G}=O(n)$, that is considering instead the orthogonal frame bundle $P\subset F$ with fibre $O(n)$. However this does not work since $P$ does not admit natural lifts of vectors on $M$. The closest related notion, the Kosmann lift~\cite{Kosmann} $v^K$ corresponding to lowering an index of~\eqref{eq:rotmatrix_explicit} and taking the antisymmetric part, does not satisfy $[v,w]^K = [v^K,w^K]$, where $v^K,w^K \in \Gs{TP/O(n)}$ are the Kosmann-lifted vectors, unless either $v$ or $w$ are  Killing (i.e. the Kosmann lift actually provides an homomorphism from the Lie algebra of infinitesimal isometries of $M$ to infinitesimal automorphisms of $P$). For a discussion see for example~\cite{LieSpinor}.

The construction that we have described so far can be easily embedded into the construction of $E$ as described in section \ref{sec:nonAb}. In particular, we can reproduce the inner product \eqref{eq:metric2gauge} and the generalised Lie derivative \eqref{eq:Bgen2gauge} after lifting to $\mathcal{A}$ and subsequently to $E$. From the identification $\tilde\Lambda=Lv$, we take  generalised vectors of the form $V = v+ Lv + \lambda$ and find for the scalar product
\begin{equation}
\label{eq:metricgravity}
   \met{V}{V'}
      = \tfrac12\iota_v \lambda' + \tfrac12\iota_{v'} \lambda
         + \tr (Lv Lv') ,
\end{equation}
while the generalised Lie derivative (or equivalently the Courant bracket) becomes
\begin{equation}
\label{eq:genLiegaugediff}
\begin{aligned}
   \Lgen_V V =
      & \comm{v}{v'} + \mathcal{L}_v\lambda'
          - \iota_{v'}\dd\lambda
          + \tfrac14 \alpha'\tr (Lv') \dd (Lv)
          + \comm{Lv}{Lv'} .
\end{aligned}
\end{equation}

Naively the $Lv$ terms in~\eqref{eq:metricgravity}  and~\eqref{eq:genLiegaugediff} look like higher-derivative corrections to the inner product and generalised Lie derivative on the conventional, unextended generalised tangent space $E'\simeq TM\oplus T^*M$. However, this interpretation is misleading precisely because $V=v+Lv+\lambda$ is really part of a larger generalised tangent space. Crucially, the one-form components $\lambda$ are patched in a way that depends non-trivially on the $\tilde{G}$-bundle patching. They are not the same objects that appear in the conventional generalised tangent space $E'$.\footnote{Of course we have the isomorphisms $E'\simeq TM\oplus T^*M$ and $E\simeq TM\oplus T^*M\oplus\adj{F}$. Thus given a choice of splitting for each bundle, that is a $B$-fields $B'$ for $E'$ and $B$ for $E$ (with different patching properties and Bianchi identities!) it is possible to map between objects in $E'$ and $E$, but there is no such natural choice.} We see that the only way to make sense of such higher-derivative corrections, both from a diffeomorphism-invariant perspective and from the patching, is to realise that they really come from the extended bundle $E$. 

One should note that working in a local coordinate frame where $\hat{e}_a=\delta_a^\mu\der_\mu$ we have $Lv^a{}_b=\der_bv^a$ and strikingly~\eqref{eq:metricgravity} and the vector and one-form part of ~\eqref{eq:genLiegaugediff} become exactly the DFT $\alpha'$-corrected expressions given in~\cite{Hohm:2013jaa}. The above construction gives a description of how to realize these $\alpha'$-corrected generalised Lie derivatives and inner products on a general curved manifold, and suggests that corrections linear in $\alpha'$ have to be of the described kind.

Another variation is to consider taking $\tilde{G}$ to be the full $O(n,n)$ group and lift not vectors $v$ but conventional generalised vectors $V=v+\lambda\in \Gs{E'}$ into $O(n,n)$ gauge transformations. This possibility was first pointed out in section 4.2 of~\cite{Bedoya:2014pma} in a DFT context. Taking $E\simeq TM\oplus T^*M\oplus \adj\tilde{F}$, where $\tilde{F}$ with fibre $O(n,n)$ is the generalised frame bundle, they defined the lifted vector $V+\Lambda\in \Gs{E}$ with $\Lambda^M{}_N = \tfrac12 (\partial_N V^M-\partial^M V_N)$. Again this is a coordinate-dependent expression, but nonetheless substituting into, for instance, the definition of the scalar product, they obtained terms like
\begin{equation}
\tfrac14(\partial_N V^M-\partial^M V_N)(\partial_M V'^N-\partial^N V'_M)=\tfrac12(\partial_N V^M-\partial^M V_N)(\partial_M V'^N)=\tfrac12 (\partial_n v^m)(\partial_m v'^n) ,
\end{equation}
where the last equality follows from the ``strong constraint'' of DFT. This term then matches the $GL(n,\bbR)$ transformation we considered and hence also gives the $\alpha'$ correction discussed in~\cite{Hohm:2013jaa}.

Again we can try and make this description properly covariant by introducing the generalised geometric analogue of the $Lv$ lift defined above.  If $\hat{U}^{AB}=\hat{E}^A\otimes\hat{E}^B$ is a basis for $\adj\tilde{F}$ defined by the generalised frame $\hat{E}_A$, then we define the lifted object
\begin{equation}
   \tilde{V} = V + LV,
\end{equation}
where $LV=(LV)_{AB}\hat{U}^{AB}$ with\footnote{Note that the form of the generalised Lie derivative means that $LV$ is actually only an element of the Lie algebra of the geometric subgroup $G_{\text{geom}}\subset O(n,n)$, corresponding to $\mathfrak{gl}(n,\bbR)$ transformations combined with $B$-shifts.}
\begin{equation}
   (LV)_{AB} = \met{\hat E_B}{ \Lgen_V \hat E_A}.
\end{equation}
One then finds, for example, that there is a generalised Lie derivative on $E$ induced from that on $E'$, given by
\begin{equation}
\label{eq:tilde-LD}
   \Lgen_{\tilde{V}}\tilde{W}
     := \widetilde{\Lgen_V W}
     = \Lgen_V W + [LV,LW]+ \met{V}{\dd (LW)}-\met{W}{\dd (LV)} .
\end{equation}
Although one still has $[\Lgen_{\tilde{V}},\Lgen_{\tilde{W}}]=\Lgen_{\Lgen_{\tilde{V}}\tilde{W}}$ and $\Lgen_{\tilde{U}}\met{\tilde{V}}{\tilde{W}}=\met{\Lgen_{\tilde{U}} \tilde{V}}{\tilde{W}}+\met{\tilde{V}}{\Lgen_{\tilde{U}} \tilde{W}}$, crucially the third condition in~\eqref{eq:CA-def} required for~\eqref{eq:tilde-LD} to define a Courant algebroid is not satisfied. This is precisely because the terms of the form $2\tr LW \dd(LV)$ are missing. This is the same problem mentioned above: there is no natural map from $E'\simeq TM\oplus T^*M$ to $E$ respecting the generalised Lie derivative structures on each. Nonetheless the existence of the lift $LV$ from $E$ to the generalised frame bundle $\tilde{F}$ is perhaps worthy of further study. In addition, it is interesting to note that the $E_{7(7)}$ generalised Lie derivative~\cite{CSW-11d} also fails to satisfy the corresponding condition, but nonetheless defines the correct structure to describe the relevant truncations of eleven-dimensional and type II supergravities.

Returning to our original reparametrisation, we again stress that the relation between gauge transformations $\tilde{\Lambda}^a{}_b=(Lv)^a{}_b$ of $\tilde G$ and diffeomorphisms does not define a subbundle of the bundle defined by~\eqref{eq:Epatch2gauge}, as it involves derivatives of the section. Note also that all the discussion in this section refers to the underlying differential structure before one introduces any of the dynamical degrees of freedom. For this reason it gives no information relating to the particular choice of  connection $\Omega^-$ that appears in the gravitational part of the Bianchi identity.

\section{General formalism}
\label{sec:gen}

Our general philosophy has been that $\alpha'$ corrections are naturally encoded by first considering further extensions of the generalised tangent space with a corresponding generalised Lie derivative, then defining a particular class of generalised connections, and finally using a Bismut-type formula to write the action.  A test of the construction would be its usefulness in ordering higher $\alpha'$ terms. The first observation is that at the very first step of defining the extended generalised tangent space there can be obstructions. To leading order in the heterotic theory these were that the Pontryagin classes of the tangent and gauge bundle cancelled. At higher $(\alpha')^n$ order, simply on dimensional ground one would expect obstructions involving classes in $n+1$ powers of the curvature. Since $\dd H$ is a four-form it is hard to envisage any higher order obstructions involving the Bianchi identity. Interestingly, at order $(\alpha')^3$ though one might though have an obstruction encoding the $B\wedge X_8$ terms.

More generally one might ask whether there is a general framework for
describing such extensions and in particular if the Bismut-type
formulae~\eqref{eq:newBismut} and~\eqref{eq:newBismut2} are generic
features. One approach is motivated by the structure of gauged
supergravity theories. Generically the gauging modifies the
supersymmetry transformations, adding new fermionic variations of the
form
\begin{equation}
\label{eq:susy-mod}
   \delta'_\epsilon\Psi = A \cdot \epsilon , \qquad \qquad
   \delta'_\epsilon\chi = B \cdot \epsilon ,
\end{equation}
where $\Psi_\mu$ are the gravitinos in the theory and $\chi$ the other
spin-$\frac12$ fermions, and $A$ and $B$ are generic matrices. There
is then a supersymmetric ``Ward identity''~\cite{s-ward,embed-lecture}
that relates these variations to the potential $V$, namely
\begin{equation}
\label{eq:sward}
   B^\dag B - A^\dag A = V \, \id ,
\end{equation}
which reminiscent of the Bismut-type formulae. These relations are
actually best described using the embedding tensor
formalism~\cite{embed}  (for a review see~\cite{embed-lecture}).

As we will now show, all generalised geometric constructions can
actually be rephrased as very particular infinite-dimensional versions
of the embedding tensor formalism. In this framework, the Bismut-type
relation follows directly from the supersymmetric Ward
identity~\eqref{eq:sward}. It also provides a formalism for addressing
how these objects might be generalised to describe higher-order corrections.
That there is a relation between generalised geometry and the
embedding tensor was already observed in~\cite{CSW-11d}, and the
formalism also played a crucial r\^ole in the reformulations given
in~\cite{EFT}, and in the discussions in~\cite{ext-geom}. Here,
though, we go a step further and show that generalised geometry can be
viewed precisely as an infinite-dimensional version of the embedding
tensor construction.

We start by recalling the basic ingredients of the formalism. One
begins with an ungauged supergravity theory in $n$ dimensions with a
global symmetry group $\mathcal{G}$ and an R-symmetry $\mathcal{H}$,
and (at least) the following content
\begin{equation}
\begin{aligned}
   \text{scalars:} & \qquad &
      M &\in  \mathcal{G}/\mathcal{H} , \\
   \text{Abelian gauge fields:} & \qquad &
      A_\mu &\in \mathcal{E} , \\
   \text{gravitinos:} & \qquad &
      \Psi_\mu & \in \mathcal{S} , \\
   \text{spin-$\tfrac12$ fields:} & \qquad &
      \chi & \in \mathcal{J} ,
\end{aligned}
\end{equation}
where $\mathcal{E}$ is a $\mathcal{G}$-representation and $\mathcal{J}$ and $\mathcal{S}$ are $\mathcal{H}$-representations. In general there may also be higher-rank $p$-form fields. The gauged theory is determined by the embedding tensor $X$. This is a map
\begin{equation}
   X : \mathcal{E} \to \adj{\mathcal{G}} ,
\end{equation}
where $\adj{\mathcal{G}}$ is the adjoint representation of $\mathcal{G}$. Supersymmetry requires that the map satisfies
\begin{align}
   \label{eq:X-Leibniz}
   (1) \quad & \comm{X(U)}{X(V)} = X(X(U)\cdot V) , \\
   \label{eq:X-rep}
   (2) \quad & \text{a particular restriction on $\mathcal{G}$-reps
         appearing in $X$} ,
\end{align}
where $X(U)\cdot V$ is the adjoint action of $X(U)$ on $V$. Note that
the first condition means that $X$ makes $\mathcal{E}$ into a Leibniz
algebra.\footnote{The relation~\eqref{eq:X-Leibniz} is more usually
   written in explicit indices: viewing $\adj{\mathcal{G}}\subset
   \mathcal{E}^*\otimes \mathcal{E}$, if elements of $\mathcal{E}$
   have components $V^M$ then the embedding tensor can be written as
   $X_{MN}{}^P$, with~\eqref{eq:X-Leibniz} taking the form
   $\comm{X_M}{X_N}=-X_{MN}{}^PX_P$.}

The tensor $X$ completely determines the gauged theory. It generates a
potential for the scalars and mass terms for the fermions, and it
describes the gauging of the kinetic terms and also the corrections to
the supersymmetry transformations. In particular, it determines the
matrices $A\in\mathcal{S}\otimes\mathcal{S}^*$ and
$B\in\mathcal{J}\otimes\mathcal{S}^*$ appearing
in~\eqref{eq:susy-mod}. In fact, for maximal supergravity theories the
representation constraint on $X$ is that, decomposing under
$\mathcal{H}$, only the $A$ and $B$ representations appear. In other
words, $A$ and $B$ are uniquely determined by $X$ and (for the maximal
case) vice versa. The potential then follows from~\eqref{eq:sward}.

Let is now see how generalised geometry falls within this
framework. For concreteness we will consider $O(d,d)\times\bbR^+$
formulation of type II theories but the same ideas work
equally well for the full $\Edd\times\bbR^+$
formulations~\cite{CSW-11d} of $d$-dimensional truncations of type II
or eleven-dimensional supergravity. The idea is to imagine
``dimensionally reducing'' the theory on a $d$-dimensional manifold
$M$ without actually truncating any of the modes, following the
original idea of de~Wit and Nicolai~\cite{dWN} (see also~\cite{GLW} in the context of generalised geometry). Thus our set of ``moduli'' is an infinite-dimensional space formed of arbitrary choices of metric, $B$-field and dilaton tensor fields on $M$, since these all transform as scalars from the point of view of the lower-dimensional theory.\footnote{Note that one is also keeping an infinite set of spin-two, and spin-$\frac{3}{2}$ fields in the lower-dimensional theory, so this framework describes the gauging of a very unconventional lower-dimensional supergravity theory.}

We can formally describe this moduli space as a coset
$\mathcal{G}/\mathcal{H}$ where we define
\begin{equation}
\begin{aligned}
   \mathcal{G} &= \text{group of diffeomorphisms and
      local $O(d,d)\times\bbR^+$ gauge transf} , \\
   \mathcal{H} &= \text{subgroup of local $O(d)\times O(d)$ gauge transf} .
\end{aligned}
\end{equation}
Mathematically we consider an (arbitrary) $O(d,d)\times\bbR^+$
principle bundle $\tilde{F}$ over $M$. The group $\mathcal{G}$ is then
the group of bundle isomorphisms, that is equivariant
diffeomorphisms, those which preserve the group action on the
fibre. Identifying the subgroup $\mathcal{H}$ is equivalent to
identifying an $O(d)\times O(d)$ sub-bundle $P\subset\tilde{F}$. The
group $\mathcal{H}$ is then the group of bundle automorphisms of
$P$. Each choice of $P\subset\tilde{E}$ is equivalent to giving the
generalised metric function $G(x)$ (more precisely this is a section of the
corresponding vector bundle), which is equivalent to specifying the
functions $g(x)$, $B(x)$ and $\phi(x)$. Thus $\mathcal{G}/\mathcal{H}$
is an infinite-dimensional space of sections.

To define $\mathcal{E}$, we start with $E$ the vector bundle over $M$
defined by the $2d$-dimensional fundamental representation of $O(d,d)$
with zero weight under the $\bbR^+$ factor. Then the space of sections
of $E$ forms a representation of $\mathcal{G}$, since any section is
mapped to another under diffeomorphisms and gauge
transformations. Thus we define $\mathcal{E}$
\begin{equation}
   \mathcal{E} = \Gs{E}, \qquad \qquad \text{space of sections of
      $E$}.
\end{equation}
Note that the $O(d,d)$ metric $\met{U}{V}$ can be thought of as
defining an equivariant map between representations of
$\mathcal{G}$. Specifically it defines a map
$\mathcal{E}\otimes\mathcal{E}\to C^\infty(M)$ where $C^\infty(M)$ is
the space of smooth functions on $M$.

It is important to note that thus far there is no requirement that
$E$ or $\tilde{F}$ have anything to do with the conventional
generalised tangent space $TM\oplus T^*M$, for the moment they are
completely general $O(d,d)\times\bbR^+$ bundles. However, let us now
turn to the embedding tensor. Remarkably, this is none other than the
generalised Lie derivative
\begin{equation}
   X(V) = \Lgen_V ,
\end{equation}
with the understanding that by definition $X$ relates $E$ to the
generalised tangent space $E\simeq TM\oplus T^*M$ and $\tilde{F}$ to
the corresponding generalised frame bundle. From the explicit
expression~\eqref{eq:genLie} we see that $\Lgen_V$ is indeed the combination of an infinitesimal diffeomorphism and an infinitesimal
$O(d,d)\times\bbR^+$ gauge transformation as required. From the
Leibniz property~\eqref{eq:relLieCourant} we see that the
condition~\eqref{eq:X-Leibniz} is identically satisfied.

The representation constraint connects $X$ to modifications to the supersymmetry variations~\eqref{eq:susy-mod}. For $O(d,d)\times\bbR^+$ generalised geometry the spin-$\frac12$ fermions $\chi$ in the ``dimensionally reduced'' theory come from the internal components of the gravitinos $\psi^\pm_m$, while the variation of the non-compact gravitinos  $\Psi_\mu$ actually gets related to the variation of the internal dilatinos $\rho^\pm$. Thus we can identify the corresponding infinite-dimensional spaces
\begin{equation}
   \mathcal{S}^\pm = \Gs{S(C_\pm)} , \qquad \qquad
   \mathcal{J}^\pm = \Gs{C_\mp\otimes S(C_\pm)} .
\end{equation}
If $|\vol_G|$ is the density defined by the generalised metric $G$, then we have the inner products on $\mathcal{S}^\pm$ and $\mathcal{J}^\pm$
\begin{equation}
\label{eq:SJnorm}
   \Met{\epsilon}{\epsilon^{\prime\pm}}
      = \int_M \norm{\vol_G} \,
         \bar{\epsilon}^\pm\epsilon^{\prime\pm} ,
   \qquad \qquad
   \Met{\psi^\pm}{\psi^{\prime\pm}}
      = \int_M \norm{\vol_G} \,
         \bar{\psi}^\pm\cdot\psi^{\prime\pm} ,
\end{equation}
where $\bar{\psi}^+\cdot\psi^{\prime+}
=\bar{\psi}^{+\bar{a}}\psi^{\prime+}_{\bar{a}}$ and $\bar{\psi}^-\cdot\psi^{\prime-}
=\bar{\psi}^{-a}\psi^{\prime-}_{a}$. From the generalised geometry construction we can then read off the operators $A$ and $B$ that appear in~\eqref{eq:susy-mod}. We find
\begin{equation}
\begin{aligned}
   A^+ &= \gamma^aD_a , & & \qquad \qquad &
   B_{\bar{a}}^+ &= D_{\bar{a}} , & \\
   A^- &= \gamma^{\bar{a}}D_{\bar{a}} , & &&
   B_a^- &= D_a .
\end{aligned}
\end{equation}
Thus these four unique operators can be viewed as the decomposition of $X=\Lgen$ into $\mathcal{H}$ representations. To see this one replaces $\Lgen_V=\Lgen^D_V$, which is possible because $D$ is torsion-free, and then, because $D$ is compatible with the generalised metric, one can decompose into $\mathcal{H}$ representations. Put another way, given the data of $\Lgen$ (to define a notion of torsion) and $G$, we know that the operators $A^\pm$ and $B^\pm$ are unique. Finally, in the type II theory we have the two Bismut relations~\cite{CSW1}
\begin{equation}
\label{eq:newBismut4}
\begin{aligned}
   D^aD_a\epsilon^-
      - \gamma^{\bar{a}}D_{\bar{a}}
        \gamma^{\bar{b}}D_{\bar{b}}\epsilon^-
      = \tfrac14 S^-\epsilon^-, \\
   D^{\bar{a}}D_{\bar{a}}\epsilon^+
      - \gamma^{a}D_{a}\gamma^{b}D_{b}\epsilon^+
      = \tfrac14 S^+\epsilon^+,
\end{aligned}
\end{equation}
where $S^+=S^-$ is the Lagrangian for the NSNS fields. We see that these are precisely the supersymmetric Ward identities~\eqref{eq:sward}, with the appropriate conjugate operators $(A^{\pm})^\dag$ and $(B^{\pm})^\dag$ defined using the norms~\eqref{eq:SJnorm}.

Formally, one can consider the full ten-dimensional theory as a reduction to zero dimensions on a ten-dimensional manifold $M$. The ungauged theory is completely trivial since there are no kinetic terms in zero dimensions. The whole theory appears via the gauging, so that, for example, the ``potential on the internal space'' actually gives the full NSNS Lagrangian. But this immediately raises the point that this gives a general formalism for describing general supersymmetric theories, and in particular the $\alpha'$ corrections. In fact, the description of the heterotic theory given in section~\ref{sec:het-corr} is exactly of this type. The corrections appeared as extension of the generalised tangent space (and how the coset structure appeared). One might, of course, also consider modifications of the embedding tensor -- that is, in the this context the generalised Lie derivative. This might be the way one can relate to the construction from the previous section.

\section{Comment on NS5-branes}
\label{sec:NS5}

We would like to conclude with some brief comments about how NS5-brane sources might be treated within the generalised geometry framework.

The modification of the Bianchi identity due to the fivebranes is another instance where $H$ is no longer a curvature of a connection on a gerbe.\footnote{Indeed, as explained in \cite{fhmm} the global treatment of the branes requires nontrivial transition functions that are a combination of antisymmetric tensor gauge transformations and diffeomorphisms. } In the presence of fivebranes the Bianchi identity \eqref{eq:hetBianchi*} changes to
\begin{equation}
\label{eq:NS5}
   \dd H = \tfrac{1}{4} \alpha' \left[\tr F\wedge F  -
      \tr R(\Omega^-)\wedge R(\Omega^-) \right]+ 2\pi  \delta ,
\end{equation}
where $\delta$ is a four-form which integrates to one in the directions transverse to the fivebrane and has delta function support on the fivebrane. This means in particular that in compact geometries one does not need to balance the gravitational contribution to~\eqref{eq:NS5} by instantons only but may use NS5-branes. Topologically we can think of $\delta$ as representing the cohomology class $[W]$ that is Poincar\'e dual to the collection $W=W_1+\dots + W_n$ of the individual cycles $W_i$ on which each  fivebrane is wrapped. In cohomology the Bianchi identity implies
\begin{equation}
\label{eq:NS5int}
   \tfrac{1}{2}p_1(M) = c_2 + [W] ,
\end{equation}
where $c_2$ is the second Chern class of the gauge bundle.

A central point of the discussion in this paper is that the standard heterotic Bianchi identity is precisely captured by the generalised geometry on the extended tangent space $E$. In particular, it is encoded in the transitive Courant algebroid structure defined by the generalised Lie derivative. It is then very interesting to understand if the same structures can be used to encode the fivebrane terms as well. One might simply add an additional auxiliary gauge bundle with second Chern class equal to $[W]$. This would at least capture the correct topology, but one then requires  local data that there are no corresponding gauge fields, and a singular delta function source.

There are two different simple situations were aspects of the source contribution to \eqref{eq:NS5} are particularly interesting. The first is when the fivebrane embedding (i.e. the normal bundle) is nontrivial. The local $n$-dimensional geometry near the brane can be viewed as a transverse space given by a ball with a sphere boundary $S^{n-7}$ fibered over the six-dimensional world volume $W$ of the fivebrane. The second case is when the manifold is a product $M\times K$ such that $K$ is a compact space transverse to the fivebranes. The source contribution is cohomologically equivalent to the (non-trivial) volume element on $K$. In both cases, one can consider decomposing the $SO(n)$ symmetry to $SO(6) \times SO(n-6)$.  This breaking can introduce some novelties into our previous construction. We shall not try to address the general case but consider the simplest interesting example where $K$ is a $K3$ compactification of heterotic string. The fivebranes are points on $K$ and have a trivial normal bundle. It is well-known that, while in absence of the branes, the cohomology condition~\eqref{eq:NS5int} for the $K3$ background requires the instanton number for the internal part of the gauge field to be 24, in the presence of $Q_5$ fivebranes the required instanton number becomes $24-Q_5$.

We can then make a minimal identification of the relevant gauge bundles and obstructions involved. Over $K$ we have a non-trivial gauge theory with group $H\subset E_8 \times E_8$ or $SO(32)$ and corresponding bundle $P_H$. We also have the frame bundle which reduces to an $\SU(2)$-bundle $P_{\SU(2)}$ since $K3$ has $\SU(2)$ holonomy. Over $M$, there is a bundle $P_G$ for the ``unbroken'' part of the gauge group (that is the commutant of $H$ in $E_8$ or $SO(32)$) and the frame bundle $F_M$. Over $M$ the extension gives a non-trivial Bianchi identity in six dimensions, dependent on $P_G$ and $F_M$, the construction of which has already been explained. Over $K$, in the absence of fivebranes, one finds a  topologically-trivial internal part to the Bianchi identity, which depends on the $P_H$ and $P_{\SU(2)}$ bundles, and whose integrated version is the $c_2(P_H)=24$ condition. One could have come up with a more elaborate version of the construction, but in the absence of NS5-branes this minimal version captures the essential ingredients.

If we include fivebranes we have a new ingredient, a four-form $V(K) \simeq \Lambda^4 T^*K$, cohomologically equivalent to a volume form on $K3$ and denoting the fivebrane charge. The construction over $M$ is unchanged, but we now have an obstruction on $K3$, given by cohomology condition~\eqref{eq:NS5int} which integrates to $c_2(P_G) + Q_5 = 24$ and the generalised tangent space and generalised Lie derivative on $K$ is determined by $P_H$, $P_{\SU(2)}$ and $V(K)$. One possibility is to represent $V(K)$ by another bundle $P_5$ with $c_2(P_5)=[V(K)]$, but somehow restrict this to be non-dynamic and have a localised field strength. Alternatively, one could relax the third condition~\eqref{eq:CA-def} so that the Courant algebroid is no longer transitive, and the original Bianchi identity need not hold. However, this relaxation must be in a controlled way that depends on the class $[V(K)]$. One might imagine for example further extending the generalised tangent space by $\Lambda^4T^*K$, however $V(K)$ plays the role of charge rather than a gauge parameter and so this is not the most natural approach to take. Geometrically the quantised fivebrane is captured by a sort of singular 2-gerbe (the charge is a four-form rather than a three-form) perhaps best described using sheaves.

The discussion here is rather sketchy and speculative, and it would be interesting to have a more complete description of NS5-branes in generalised geometry, notably away from the small-instanton limit. It also makes no attempt to capture aspects of the theory on the fivebrane itself. It may also be of some interest to work out a more complete description of six-dimensional $(0,1)$ theories and the higher-derivative $\alpha'$ couplings in this formalism.


\section*{Acknowledgments}


We would like to thank Mariana Gra\~na for collaboration at the initial stage of the project, and  Marco Gualtieri for useful discussions. This work was supported in part by the ERC Starting Grant 259133 -- ObservableString, the German Research Foundation DFG within the Cluster of Excellence ``QUEST'' (AC), the Agence Nationale de la Recherche under the grant 12-BS05-003-01 (RM), the EPSRC Programme Grant ``New Geometric Structures  from String Theory'' EP/K034456/1 (DW) and the STFC Consolidated Grant ST/J0003533/1 (DW). AC also thanks CEA Saclay for hospitality during the completion of this work.
\appendix

\section{Type II theories}
\label{sec:typeII}

In section~\ref{sec:het} we developed the formalism for a generalised tangent bundle with structure group $O(n)_+\times G_1\times G_2\times O(n)_-$, and proceeded to show that by identifying $G_2$ with the $O(n)_+$ we obtain the supergravity limit of the heterotic string to first order in $\alpha'$. One might then wonder what would happen if we also identified $G_1$ with $O(n)_-$ (a similar idea has already been proposed in~\cite{dnpw}). The generalised tangent bundle $E = C_+ \oplus C_{\mathfrak{g}_1} \oplus C_{\mathfrak{g}_2} \oplus C_-$ then becomes
\begin{equation}\label{eq:E-TypeII}
   E = C_+ \oplus  \Lambda^2C_- \oplus \Lambda^2C_+ \oplus C_-,
\end{equation}
which would bring the local symmetry group down to just $O(n)_+\times O(n)_-$. This is the structure group of the generalised tangent bundle of (the NSNS sector of) type II supergravity, so does this correspond to an alternative generalised geometric description of type II? Note that the Bianchi identity~\eqref{eq:hetBianchi*}  becomes
\begin{equation}\label{eq:typeIIbianchi2}
\dd H = \tfrac14 \alpha' [\tr R(\Omega^+)\wedge R(\Omega^+) - \tr R(\Omega^-)\wedge R(\Omega^-)],
\end{equation}
but this is simply an unusual way of writing the type II Bianchi $\dd H = 0$, since we are just taking the difference of two different representatives of the same characteristic class.

The gravity multiplet of type II contains twice as many fermions, so now not only does $C_-$ have an associated spin-bundle $S(C_-)$, but we must also introduce an associated spin-bundle $S(C_+)$ to $C_+$ as well (these spin-bundles would then have to be decomposed into their different chirality components to distinguish IIA from IIB but that will not affect this discussion). Therefore equation~\eqref{eq:FermionFields} describing the representations of the fermion fields gets extended to
\begin{equation}
\begin{aligned}
\text{gravitini:} & &&\psi^-_a &\in \Gs{C_+\otimes S(C_-)},\quad &\psi^+_{\bar{a}} &\in \Gs{C_-\otimes S(C_+)},\\
\text{``gaugini'':} & && \zeta^-_{aa'} &\in \Gs{\Lambda^2C_+ \otimes S(C_-) },\quad &\zeta^+_{\bar{a}\bar{a}'} & \in \Gs{\Lambda^2C_-\otimes S(C_+)},\\
& &&\zeta^+_{aa'} &\in \Gs{\Lambda^2C_+\otimes S(C_+) },\quad &\zeta^-_{\bar{a}\bar{a}'} &\in \Gs{\Lambda^2C_-\otimes S(C_-)},\\
\text{dilatini:}& &&\rho^- &\in \Gs{S(C_-)}, \quad &\rho^+ & \in \Gs{S(C_+)},
\end{aligned}
\end{equation}
where, as for the heterotic case, the ``gaugini'' are to be thought of as composite fields and must eventually be related to the gravitini.

Now, as in~\eqref{eq:SugraOps}, we must identify the differential operators which are constructed from generalised connections and preserve the representations of the fermion fields. These will depend on certain components of the connection which we will then demand be fixed unambiguously since we expect them to feature in the supergravity theory. They are
\begin{equation}
\begin{aligned}
\label{eq:SugraOpsII}
&\gamma^{\bar{a}}D_{\bar{a}}\lambda^-, \quad & D^a\varphi^-_a, \quad & D^{aa'}\xi^-_{aa'}, \quad & D^{\bar{a}\bar{a}'}\xi^-_{\bar{a}\bar{a}'},\quad &\in \Gs{S(C_-)},\\
&\phantom{\gamma^{\bar{a}}D_{\bar{a}}\lambda^-,} \quad & \phantom{D^a\varphi^-_a,} \quad & D_a \lambda^-, \quad  & \gamma^{\bar{a}}D_{\bar{a}}\varphi^-_a, \quad &\in \Gs{C_+\otimes S(C_-)},\\
&\phantom{\gamma^{\bar{a}}D_{\bar{a}}\lambda^-,} \quad & \phantom{D^a\varphi^-_a,} \quad& D_{aa'}\lambda^-, \quad & \gamma^{\bar{a}}D_{\bar{a}}\xi^-_{aa'} \quad &\in \Gs{\Lambda^2C_+\otimes S(C_-)},\\
&\phantom{\gamma^{\bar{a}}D_{\bar{a}}\lambda^-,} \quad & \phantom{D^a\varphi^-_a,} \quad& D_{\bar{a}\bar{a}'}\lambda^-, \quad & \gamma^{\bar{b}}D_{\bar{b}}\xi^-_{\bar{a}\bar{a}'} \quad &\in \Gs{\Lambda^2C_-\otimes S(C_-)},\\
& \gamma^{\bar{a}}D_{\bar{a}}\lambda^+, \quad & D^{\bar{a}}\varphi^+_{\bar{a}}, \quad & D^{\bar{a}\bar{a}'}\xi^+_{\bar{a}\bar{a}'}, \quad & D^{aa'}\xi^+_{aa'},\quad &\in \Gs{S(C_+)},\\
&\phantom{\gamma^{\bar{a}}D_{\bar{a}}\lambda^-,} \quad & \phantom{D^a\varphi^-_a,} \quad& D_{\bar{a}} \lambda^+, \quad  & \gamma^{a}D_{a}\varphi^+_{\bar{a}}, \quad &\in \Gs{C_-\otimes S(C_+)},\\
&\phantom{\gamma^{\bar{a}}D_{\bar{a}}\lambda^-,} \quad & \phantom{D^a\varphi^-_a,} \quad& D_{\bar{a}\bar{a}'}\lambda^+, \quad & \gamma^{a}D_{a}\xi^+_{\bar{a}\bar{a}'} \quad &\in \Gs{\Lambda^2C_-\otimes S(C_+)},\\
&\phantom{\gamma^{\bar{a}}D_{\bar{a}}\lambda^-,} \quad & \phantom{D^a\varphi^-_a,} \quad& D_{aa'}\lambda^+, \quad & \gamma^{b}D_{b}\xi^+_{aa'} \quad &\in \Gs{\Lambda^2C_+\otimes S(C_+)}.
\end{aligned}
\end{equation}
It is now clear that we have a far more constrained system than in the heterotic case. In fact, if, following~\eqref{eq:SugraOps}, we try to solve for a generalised connection compatible with the reduced structure, we find that we no longer have the necessary freedom to consistently make the identification~\eqref{eq:E-TypeII}.

To see this explicitly, let us solve the constraints~\eqref{eq:SugraOpsII} but not yet impose compatibility with the reduced $O(n)_+\times O(n)_-$ structure, i.e., let us just require that the connection is $O(n)_+\times G_1\times G_2\times O(n)_-$. Then we obtain the solution
\begin{subequations}\label{eq:OddddLC}
\begin{align}
    D_a w_+^b
       &= \nabla_a w_+^b - \tfrac{1}{6}H_{a}{}^{b}{}_{c}w_+^{c}
           - \tfrac{2}{n-1}\big(
              \delta_a{}^b \partial_c\phi-\eta_{ac}\partial^b\phi \big)w_+^c
           + Q_a{}^b{}_c w_+^c  ,\label{eq:OddddLC12} \\
     D_{\alpha^i} w_+^{b} &= \begin{cases}
       \tfrac{1}{2}\sqrt{\alpha'}F^{1b}{}_{a}{}_{\alpha^1}
           w_+^{a} & \text{if $i=1$} \\
       -\tfrac{1}{2}\sqrt{\alpha'}F^{2b}{}_{a}{}_{\alpha^2}
           w_+^{a} & \text{if $i=2$}
           \end{cases} , \label{eq:OddddLC22}\\
    D_{\bar{a}} w_+^b &= \nabla_{\bar{a}} w_+^b -
       \tfrac{1}{2}H_{\bar{a}}{}^b{}_cw_+^c,
       \label{eq:OddddLC32}\\
    D_a  w_{\mathfrak{g}_i}^{\beta^i}
       &= \nabla^i_a w_{\mathfrak{g}_i}^{\beta^i} ,
       \label{eq:OddddLC42}\\
    D_{\alpha^i}  w_{\mathfrak{g}_j}^{\beta^j} &=
      Q_{\alpha^i}{}^{\beta^j}{}_{\gamma^j}w_{\mathfrak{g}_j}^{\gamma^j},
      \label{eq:OddddLC52}\\
    D_{\bar{a}} w_{\mathfrak{g}_i}^{\beta^i} &=
      \nabla^i_{\bar{a}} w_{\mathfrak{g}_i}^{\beta^i},
      \label{eq:OddddLC62}\\
    D_a w_-^{\bar{b}}
       &= \nabla_a w_-^{\bar{b}}
       + \tfrac{1}{2}H_a{}^{\bar{b}}{}_{\bar{c}}w_-^{\bar{c}} ,
       \label{eq:OddddLC72}\\
    D_{\alpha^i} w_-^{\bar{b}} &= \begin{cases}
       -\tfrac{1}{2}\sqrt{\alpha'}F^{1\bar{b}}{}_{\bar{a}}{}_{\alpha^1}
           w_-^{\bar{a}} & \text{if $i=1$} \\
       \tfrac{1}{2}\sqrt{\alpha'}F^{2\bar{b}}{}_{\bar{a}}{}_{\alpha^2}
           w_-^{\bar{a}} & \text{if $i=2$}
           \end{cases} , \label{eq:OddddLC82}\\
    D_{\bar{a}} w_-^{\bar{b}} &= \nabla_{\bar{a}} w_-^{\bar{b}}
       + \tfrac{1}{6}H_{\bar{a}}{}^{\bar{b}}{}_{\bar{c}}w_-^{\bar{c}}
       - \tfrac{2}{n-1}\big(
          \delta_{\bar{a}}{}^{\bar{b}} \partial_{\bar{c}}\phi
          - \eta_{\bar{a}\bar{c}}\partial^{\bar{b}}\phi \big)w_-^{\bar{c}}
       + Q_{\bar{a}}{}^{\bar{b}}{}_{\bar{c}} w_-^{\bar{c}} ,\label{eq:OddddLC92}
\end{align}
\end{subequations}
where
\begin{equation*}\begin{aligned}
Q_{[abc]}=0, \quad Q_{[\bar{a}\bar{b}\bar{c}]}=0,\quad Q_{[\alpha^i,\beta^i,\gamma^i]}=0,\\
Q_{a}{}^a{}_b = 0,\quad Q_{\bar{a}}{}^{\bar{a}}{}_{\bar{b}}=0,\quad Q_{\alpha^i}{}^{\alpha^i}{}_{\beta^i}=0.
\end{aligned}\end{equation*}
This should be compared with~\eqref{eq:OdggdLC}. Note that far more components get fixed, and in particular equations~\eqref{eq:OddddLC42} and~\eqref{eq:OddddLC62} no longer have any dependence on the unconstrained $Q$ tensors. So if we were to perform the identification~\eqref{eq:E-TypeII} so that
\begin{equation}
\alpha_1 = [\bar{a}\bar{a}'],\quad \alpha_2 = [aa'],
\end{equation}
we find that there is no solution for a compatible connection -- say we identify one of the equations~\eqref{eq:OddddLC42}, corresponding to
\begin{equation}
D_a w_-^{\bar{b}\bar{b}'} = \partial_a w_-^{\bar{b}\bar{b}'} + A^1_a{}^{\bar{b}}{}_{\bar{b}}w_-^{\bar{c}\bar{b}'}+ A^1_a{}^{\bar{b}'}{}_{\bar{b}}w_-^{\bar{b}\bar{c}},
\end{equation}
with~\eqref{eq:OddddLC72},
\begin{equation}
D_a w_-^{\bar{b}} = \partial_a w_-^{\bar{b}} + \Omega^+_a{}^{\bar{b}}{}_{\bar{c}}w_-^{\bar{c}}.
\end{equation}
This would mean taking
\begin{equation}
A^1 = \Omega^+ = \omega^{\text{LC}}+\tfrac12 H.
\end{equation}
But then we have a contradiction, since~\eqref{eq:OddddLC62} now reads
\begin{equation}\begin{aligned}
D_{\bar{a}} w_-^{\bar{b}\bar{b}'} &= \partial_{\bar{a}} w_-^{\bar{b}\bar{b}'} + A^1_{\bar{a}}{}^{\bar{b}}{}_{\bar{b}}w_-^{\bar{c}\bar{b}'}+ A^1_{\bar{a}}{}^{\bar{b}'}{}_{\bar{b}}w_-^{\bar{b}\bar{c}},\\
&= \nabla_{\bar{a}} w_-^{\bar{b}\bar{b}'}+\tfrac12 H_{\bar{a}}{}^{\bar{b}}{}_{\bar{c}} w_-^{\bar{c}\bar{b}'}+\tfrac12 H_{\bar{a}}{}^{\bar{b}'}{}_{\bar{c}} w_-^{\bar{b}\bar{c}},
\end{aligned}\end{equation}
which is inconsistent with~\eqref{eq:OddddLC92}
\begin{equation}
D_{\bar{a}} w_-^{\bar{b}} = \nabla_{\bar{a}} w_-^{\bar{b}}
       + \tfrac{1}{6}H_{\bar{a}}{}^{\bar{b}}{}_{\bar{c}}w_-^{\bar{c}}
       - \tfrac{2}{n-1}\big(
          \delta_{\bar{a}}{}^{\bar{b}} \partial_{\bar{c}}\phi
          - \eta_{\bar{a}\bar{c}}\partial^{\bar{b}}\phi \big)w_-^{\bar{c}}
       + Q_{\bar{a}}{}^{\bar{b}}{}_{\bar{c}} w_-^{\bar{c}},
\end{equation}
for any $Q$. We are forced to conclude that there is no solution for a generalised connection which is both compatible with the structure resulting from the identification $G_2 \rightarrow O(n)_+$ and $G_1 \rightarrow O(n)_-$ and which satisfies the supergravity constraints~\eqref{eq:SugraOpsII} (unless $H=\dd\phi=0$).

If we nevertheless press on without making the identification, we could then try to use solution~\eqref{eq:OddddLC} to construct an action, using the Bismut-type formula~\eqref{eq:newBismut}. However, since we now have two spin-bundles $S(C_{\pm})$ we actually have two equally valid equations defining a scalar
\begin{equation}
\label{eq:newBismut3}
\begin{aligned}
\gamma^{\bar{a}}D_{\bar{a}}\gamma^{\bar{b}}D_{\bar{b}}\epsilon^- - D^aD_a\epsilon^--D^{\alpha^1}D_{\alpha^1}\epsilon^-
+D^{\alpha^2}D_{\alpha^2}\epsilon^-=
-\tfrac14 S^-\epsilon^-,
\\
\gamma^{a}D_{a}\gamma^{b}D_{b}\epsilon^+ - D^{\bar{a}}D_{\bar{a}}\epsilon^+
+D^{\alpha^1}D_{\alpha^1}\epsilon^+
-D^{\alpha^2}D_{\alpha^2}\epsilon^+
=-\tfrac14 S^+\epsilon^+,
\end{aligned}
\end{equation}
where the signs in the ``gaugino'' terms were chosen so that right-hand side is just a scalar (a necessary requirement for a supersymmetric action) -- in general there would also be four-form terms, but with this sign choice these turn out to just be the Bianchi for $H$, which vanishes identically. We have thus defined two a priori different scalars
\begin{equation}
\label{eq:BosonicAction23}
S^{\mp} = s +4 \nabla^2 \phi -4 (\partial \phi)^2
	- \tfrac{1}{12} H^2 \mp\tfrac{\alpha'}{8}\tr_{1} (F^{1})^2\pm\tfrac{\alpha'}{8}\tr_{2} (F^{2})^2.
\end{equation}
The only way we can consistently choose an action then is if these happen to coincide, which would imply that $\tr_{1} (F^{1})^2=\tr_{2} (F^{2})^2$.

This would in fact be the case for type II, since the identity is satisfied if $A^1=\Omega^+$ and $A^2=\Omega^-$, but as we saw already we cannot make such a choice to begin with unless $H=\dd\phi=0$ (for which $S^{\pm}=s$ would also agree trivially).  We conclude that type II theories  (nor any theory involving two spin-bundles and a non-trivial gauge bundle) do not admit corrections linear in $\alpha'$.

Note that we have so far ignored the RR fields. One might wonder if they can be described by extending the conventional generalised tangent space by the Clifford algebra rather than a Lie algebra, and interpreting the resulting $\tfrac{1}{8}\alpha'\tr F^2$ terms
as kinetic terms for the  RR fields. However, at least the most naive such extension gives a different RR gauge algebra from that appearing in the supergravity.

\end{document}